\documentclass[10pt,english,journal]{IEEEtran}

\usepackage{amsmath,amssymb,amsfonts,mathtools}
\usepackage{amsthm}
\usepackage{bm}
\usepackage{booktabs}
\usepackage{algorithm}
\usepackage[noend]{algpseudocode}
\usepackage{graphicx}
\usepackage{microtype}
\usepackage[hidelinks]{hyperref}
\usepackage[nameinlink]{cleveref}
\crefname{table}{Table}{Tables}
\usepackage{url}
\usepackage{balance}
\usepackage{cite}
\usepackage{enumitem}
\usepackage{siunitx}
\usepackage{verbatim}
\usepackage{caption}
\usepackage{mdframed}
\usepackage[utf8]{inputenc}
\DeclareUnicodeCharacter{2009}{\,}   
\DeclareUnicodeCharacter{2013}{--}   

\newcommand{\R}{\mathbb{R}}
\newcommand{\E}{\mathbb{E}}

\newcommand{\norm}[1]{\left\lVert #1 \right\rVert}
\newcommand{\abs}[1]{\left\lvert #1 \right\rvert}
\newcommand{\ip}[2]{\left\langle #1,#2 \right\rangle}
\DeclareMathOperator{\diag}{diag}
\DeclareMathOperator{\Tr}{Tr}
\DeclareMathOperator{\Var}{Var}

\newcommand{\bigO}{\mathcal{O}}
\newcommand{\Cov}{\mathrm{Cov}}

\newcommand{\N}{\mathcal{N}}
\newcommand{\KL}{\mathrm{KL}}

\theoremstyle{plain}

\theoremstyle{remark}

\usepackage{tabularx}   
\usepackage{array}      
\allowdisplaybreaks 
\newcommand{\PP}{\mathbb{P}}
\newcommand{\cF}{\mathcal{F}}

\newcommand{\X}{\mathbf{X}}
\newcommand{\bx}{\mathbf{x}}
\newcommand{\by}{\mathbf{y}}
\newcommand{\bs}{\mathbf{s}}

\newcommand{\ct}{\mathrm{ct}}
\newcommand{\SSM}{\mathrm{SSM}}
\newcommand{\cN}{\mathcal{N}}

\usepackage{hyphenat}

\newcommand{\bmu}{\boldsymbol{\mu}}
\newcommand{\bd}{\boldsymbol{d}}
\newcommand{\bsig}{\boldsymbol{\sigma}}

\usepackage{bbm}

\usepackage{nicefrac}

\newcommand{\Dec}{\mathrm{Dec}}

\usepackage{float} 

\newcommand{\Act}{A_{\mathrm{ct}}}
\newcommand{\Bref}{B_{\mathrm{ref}}}
\newcommand{\leg}[1]{(2#1+1)}
\newcommand{\sqleg}[2]{\sqrt{\leg{#1}\,\leg{#2}}}

\newcommand{\bepsilon}{\boldsymbol{\epsilon}}

\theoremstyle{definition}

\usepackage{adjustbox}
\usepackage{comment}

\newcommand{\bz}{\boldsymbol{z}}
\newcommand{\bb}{\boldsymbol{b}}
\usepackage[table]{xcolor}

\newcommand{\bsigma}{\boldsymbol{\sigma}}
\newcommand{\WMSection}[1]{%
  \Statex \vspace{2pt}%
  \noindent\colorbox{yellow!20}{%
    \parbox{\dimexpr\linewidth-2\fboxsep\relax}{\textbf{#1}}%
  }%
  \vspace{2pt}%
}

\makeatletter
\renewcommand{\SSM}{\ifmmode\mathrm{SSM}\else\textsc{SSM}\fi}
\makeatother

\title{Agentic World Modeling for 6G: Near-Real-Time Generative State-Space Reasoning}

\author{
Farhad~Rezazadeh,~\IEEEmembership{Member,~IEEE},
Amir~Ashtari~Gargari,~\IEEEmembership{Member,~IEEE},
Hatim~Chergui,~\IEEEmembership{Senior~Member,~IEEE},
Sandra~Lag\'en,~\IEEEmembership{Senior~Member,~IEEE},
Merouane~Debbah,~\IEEEmembership{Fellow,~IEEE}, Houbing~Song,~\IEEEmembership{Fellow,~IEEE},~and~Lingjia~Liu,~\IEEEmembership{Fellow,~IEEE}

\IEEEcompsocitemizethanks{\IEEEcompsocthanksitem F. Rezazadeh is with the BrainOmega and the Technical University of Catalonia (UPC), 08028 Barcelona, Spain (e-mail: farhad.rezazadeh@upc.edu).}
\IEEEcompsocitemizethanks{\IEEEcompsocthanksitem A. shtari Gargari and S. Lag\'en are with the Centre Tecnologic de Telecomunicacions de Catalunya
(CTTC/CERCA), 08860 Castelldefels, Spain (e-mail: \{aashtari, slagen\}@cttc.es).}
\IEEEcompsocitemizethanks{\IEEEcompsocthanksitem H. Chergui is with i2CAT Foundation, 08034 Barcelona, Spain (e-mail: hatim.chergui@i2cat.net).}
\IEEEcompsocitemizethanks{\IEEEcompsocthanksitem M. Debbah is with the Khalifa University of Science and Technology, 127788 Abu Dhabi, UAE (e-mail: merouane.debbah@ku.ac.ae).}
\IEEEcompsocitemizethanks{\IEEEcompsocthanksitem H. Song is with the University of Maryland, Baltimore County (UMBC), 21250 Baltimore, USA (e-mail: h.song@ieee.org).}
\IEEEcompsocitemizethanks{\IEEEcompsocthanksitem L. Liu is with the Virginia Tech, 24061 Blacksburg, USA (e-mail: ljliu@vt.edu).}
}
\begin{document}
\IEEEaftertitletext{\vspace{-1.5\baselineskip}} 
\maketitle

\begin{abstract}
We argue that sixth-generation (6G) intelligence is not fluent token prediction but the capacity to \emph{imagine} and \emph{choose}---to simulate future scenarios, weigh trade-offs, and act with calibrated uncertainty. We reframe open radio access network (O-RAN) near-real-time (Near-RT) control via \emph{counterfactual dynamics} and a \emph{world modeling} (WM) paradigm that learns an action-conditioned generative state space. This enables quantitative ``what-if'' forecasting beyond large language models (LLMs) as the primary modeling primitive. Actions such as physical resource blocks (PRBs) are treated as first-class control inputs in a causal world model, and both aleatoric and epistemic uncertainty are modeled for prediction and what-if analysis. An agentic, model predictive control (MPC)-based cross-entropy method (CEM) planner operates over short horizons, using prior-mean rollouts within data-driven PRB bounds to maximize a deterministic reward. The model couples multi-scale structured state-space mixtures (MS\textsuperscript{3}M) with a compact stochastic latent to form WM--MS\textsuperscript{3}M\footnote{The source code is publicly available for non-commercial use at \url{https://github.com/frezazadeh/agentic-wm_ms3m}.}, summarizing key performance indicators (KPIs) histories and predicting next-step KPIs under hypothetical PRB sequences. On realistic O-RAN traces, WM--MS\textsuperscript{3}M cuts mean absolute error (MAE) by 1.69\% versus MS\textsuperscript{3}M with 32\% fewer parameters and similar latency, and achieves 35--80\% lower root mean squared error (RMSE) than attention/hybrid baselines with 2.3--4.1$\times$ faster inference, enabling rare-event simulation and offline policy screening.
\end{abstract}


\section{Introduction}

\IEEEPARstart{W}{ithin} the emerging paradigm of 6G agentic AI, \emph{world modeling} \cite{ha2018worldmodels, zhao2025worldmodels_cognitive_agents, zhao2025egi_worldmodels_agenticai} serves as the \emph{imagination} engine that enables autonomous 6G agents to internally simulate, reason, and anticipate network management outcomes before acting. Unlike traditional reactive agents, those equipped with world models can construct latent representations of their 6G environment, perform \emph{what-if} reasoning, and infer the causal consequences of potential actions, mirroring human-like deliberation and foresight. This capability forms the foundation for goal-directed autonomy in complex, dynamic domains such as network automation, where agents must continuously interpret telemetry, predict system behavior, and plan interventions under uncertainty. In this context, world models go beyond digital twins, which primarily replicate network state through real-time data synchronization and visualization. While digital twins reflect the network's current reality, world models learn its dynamics, allowing agents to imagine unobserved states, simulate hypothetical conditions, and optimize actions based on counterfactual reasoning. By integrating such imagination-driven reasoning, autonomous networks can progress toward TM Forum's levels 4 and 5, where networks proactively plan and negotiate actions through predictive world understanding \cite{tmforum2021autonomous}.

However, achieving such capabilities in O-RAN Near-RT control amounts to capturing, modeling, and reasoning over temporal dependencies under strict latency and safety constraints. Large language models (LLMs) are increasingly proposed for AI-native O-RAN \cite{wu2025llmxapp2, bao2025llmhric, bimo2025intentllm}, where they help with intent translation, policy/configuration generation, and natural-language interaction with operators. Yet, when LLMs are used as the \emph{primary} modeling primitive for control, several limitations emerge: 
(i) token-centric training and weak inductive bias \cite{chang2025inductivecounting} for stochastic dynamics make it difficult to learn calibrated physical behavior from KPIs; 
(ii) exposure bias \cite{bengio2015scheduled, lamb2016professor} and off-policy extrapolation \cite{fujimoto2019bcq} degrade reliability when the model is rolled out for multi-step prediction or planning; 
(iii) many LLM-like or Transformer-based pipelines silently leak future information \cite{meyer2025tsfmbench, xu2025fidelts}, which inflates offline performance but does not translate into deployable Near-RT behavior; and 
(iv) uncertainty is typically not modeled in a control-grade way, making risk assessment and safety guards difficult to design and audit. 
These gaps suggest that LLMs are better suited as \emph{orchestration and explanation layers}, while the core of Near-RT control should rely on leakage-safe, stateful, and uncertainty-aware dynamical models.

In this work, we propose an \emph{agentic world modeling} perspective that places \emph{counterfactual dynamics} \cite{liu2023cdf, venkatesh2024cwm} at the center of control. Concretely, we treat physical resource blocks (PRBs) as first-class \emph{actions} and learn a compact, causal, and probabilistic state-space model (SSM) \cite{smith2023simplified} that supports (i) accurate next-step KPI prediction with calibrated uncertainty, (ii) \emph{imagination} via action-conditioned rollouts (“what-if” forecasting), and (iii) short-horizon planning. We introduce \emph{WM--MS$^{3}$M}, which extends the strictly causal multi-scale structured state-space mixtures (MS$^{3}$M) from our previous work~\cite{Rezazadeh2025RivalingTM} by incorporating a lightweight stochastic latent variable and dual decoders. KPI windows---including PRB history---are summarized by a multi-scale SSM front end; a diagonal-Gaussian latent \cite{rezende2014stochastic} captures residual uncertainty; and decoders produce both a full next-step frame and a heteroscedastic \cite{kendall2017uncertainties} target head. A model predictive control (MPC) planner \cite{rawlings2017mpcbook} uses the cross-entropy method (CEM) \cite{rubinstein1999cem, walker2025dscemmpc} in the model's standardized space, operating within a data-driven PRB admissible set to avoid out-of-distribution actions. This design deliberately separates \emph{forecasting} from \emph{decision making}. The world model remains auditable and non-decisional, exposing predictive means, quantiles, and risk summaries, while the planner optimizes a short-horizon reward trading signal quality (signal-to-interference-plus-noise ratio (SINR)/spectral efficiency (SE)/reference signal received power (RSRP) against block error rate (BLER)/Delay/PRB usage with a smoothness term. The same substrate supports imagination for proactive what-if analysis and policy evaluation without online trials.

\subsection{Related Work}
\label{sec:related-work-summary}

\subsubsection{LLM-based and Agentic O-RAN Analytics}
Recent works explore LLM-centric and agentic paradigms for AI-native O-RAN. Intent-based management, xApp/rApp orchestration, and human-in-the-loop automation via LLMs have been proposed in \cite{wu2025llmxapp2, bao2025llmhric, bimo2025intentllm}, where LLMs translate operator intents into policies or configurations and help coordinate network functions. These approaches emphasize flexibility and ease of interaction but generally treat the underlying network behavior as a black box and focus on high-level control or policy templates. They do not provide a calibrated, action-conditioned dynamical model of KPIs, nor do they explicitly support quantitative counterfactual analysis under resource constraints.

\subsubsection{Efficient Sequence Models and Time-Series Foundations}
A large body of work has improved sequence modeling efficiency and expressiveness for long horizons. RWKV \cite{peng2023rwkv} blends RNN-style recurrence with attention-like mixing for streaming inference, but remains sensitive to exposure bias under one-step-ahead evaluation. Performers \cite{choromanski2021rethinking} replace softmax attention with kernelized approximations, achieving linear-time attention suitable for long contexts, while RetNet \cite{sun2023retentive} introduces retentive state-space updates with $O(T)$ complexity and strong long-memory capabilities. Time-series foundation models such as Chronos-GPT and Chronos-T5 \cite{ansari2024chronos} tokenize numerical sequences and leverage language-modeling priors for forecasting, with Chronos-GPT using a causal decoder and Chronos-T5 using a denoising seq2seq architecture with bidirectional context. These models provide powerful priors for generic sequence prediction, but: (i) they are typically trained and evaluated with token-centric losses that are not unit-aware; (ii) many use bidirectional context or sequence-level normalization that would amount to data leakage in Near-RT control; and (iii) they rarely treat control variables (e.g., PRBs) as explicit \emph{actions} in a causal dynamical system.

\subsubsection{State-Space Models and World Models for Control}
Modern SSMs such as S4 and variants \cite{gu2020hippo} provide structured, long-memory filters with stable dynamics and linear-time convolutions. These architectures have been successfully applied to generic sequence tasks and time series, and they naturally align with the controlled dynamical-systems view of wireless networks. In parallel, world models and model-based RL (e.g., \cite{ha2018worldmodels,hafner2020dreamer,hafner2021dreamerv2,sutton1991dyna}) show that learned latent dynamics can support imagination, planning, and sample-efficient decision making via MPC or policy optimization over simulated rollouts. However, most existing applications of SSMs to networking focus on observational forecasting (e.g., traffic/load prediction) without explicit control channels or counterfactual planning, and most world-modeling work targets robotics or games with different observables and constraints than O-RAN telemetry. Recent work on counterfactual or causal world models \cite{liu2023cdf, venkatesh2024cwm} moves closer to our goal but does not address Near-RT RIC integration, PRB-constrained planning, or leakage-safe telemetry pipelines.

\subsection{Contributions}
\label{subsec:contributions}

\noindent This paper makes the following contributions, each directly tied to the gaps above:

\begin{itemize}[leftmargin=*,topsep=1pt,itemsep=2pt]

\item \textbf{Paradigm: Agentic world modeling for 6G (prediction/imagination $\rightarrow$ choice).} 
We formalize a world-model-first contract for O-RAN in which a \emph{calibrated, action-conditioned} model provides both factual prediction and counterfactual imagination, and an outer-loop planner (MPC/CEM) chooses actions. 
\emph{Gap addressed:} Unlike LLM-centric or rule-based approaches \cite{wu2025llmxapp2, bao2025llmhric, bimo2025intentllm}, which entangle modeling and policy in token space, our design makes PRBs explicit actions and separates prediction from decision, improving safety, auditability, and robustness to data shifts.

\item \textbf{Architecture: WM--MS\textsuperscript{3}M (multi-scale SSM + compact latent + dual decoders).} 
We extend the strictly causal MS\textsuperscript{3}M backbone \cite{Rezazadeh2025RivalingTM} with a lightweight stochastic latent and two coordinated decoders (a full-frame head and a heteroscedastic target head with a bounded AR skip). 
\emph{Gap addressed:} Existing SSM-based or foundation time series models \cite{gu2020hippo, ansari2024chronos} either ignore uncertainty or treat control variables as passive features. Our architecture provides calibrated epistemic/aleatoric uncertainty \cite{kendall2017uncertainties} and explicitly conditions on PRBs, enabling control-grade forecasting and what-if rollouts.

\item \textbf{Leakage-safe, unit-aware learning pipeline for O-RAN telemetry.} 
We enforce chronological splits, train-only scalers, KL annealing \cite{Cui2025GKL}, posterior/prior mixing, and light robustness augmentations (feature-channel dropout, small input noise). 
\emph{Gap addressed:} Many sequence models implicitly leak future information via bidirectional context or global normalization, which is incompatible with Near-RT deployment. Our pipeline ensures that offline scores reflect true on-policy behavior with KPIs reported in physical units.

\item \textbf{Quantified uncertainty and action-conditioned what-if forecasting.} 
Test-time prior sampling yields means, intervals, and risk summaries for next-step and short-horizon KPIs under candidate PRB sequences constrained to data-driven admissible ranges. 
\emph{Gap addressed:} Traditional predictors provide point estimates only, and black-box policies cannot explain risk. WM--MS\textsuperscript{3}M enables questions like “If we $\pm 20\%$ PRBs, what is the BLER exceedance risk in the next $H$ steps?”, supporting risk-aware control and offline policy screening.

\item \textbf{Short-horizon planning via MPC/CEM in standardized space.}
We instantiate a Near-RT-compatible planner that optimizes a shaped reward trading SINR/SE/RSRP against BLER/Delay and PRB cost with a smoothness penalty, all in standardized space, and maps actions back to physical units. 
\emph{Gap addressed:} Prior O-RAN control strategies are often heuristic and myopic, while generic RL can be sample-inefficient or unsafe. Our MPC/CEM planner is sample-efficient, auditable, and constrained to in-distribution PRB ranges derived from training data.

\item \textbf{Accuracy–efficiency under Near-RT constraints.}
On realistic O-RAN traces, WM--MS\textsuperscript{3}M improves MAE by \SI{1.69}{\percent} over an equally causal MS\textsuperscript{3}M while using $\approx\SI{32}{\percent}$ fewer parameters and similar latency, and outperforms attention/hybrid baselines by $35\text{–}80\%$ RMSE with $2.3\text{–}4.1\times$ faster inference. 
\emph{Gap addressed:} This demonstrates that a world model with SSM inductive bias and compact latent can deliver control-grade accuracy at edge-friendly cost, addressing the practical deployability gap of many Transformer-family models.

\item \textbf{Reproducible artifacts and O-RAN integration.}
We release code for WM--MS\textsuperscript{3}M, the leakage-safe training/inference pipeline, and MPC/CEM planning (non-commercial license), plus curated O-RAN KPI resources, and we explicitly map the world model and planner into the Near-RT RIC/xApp workflow. 
\emph{Gap addressed:} This bridges the frequent disconnect between algorithmic proposals and O-RAN-compliant implementations, enabling follow-on work on agentic world models in 6G networks.
\\
\end{itemize}

The remainder of this paper is organized as follows.
\Cref{sec:system-model} details the system model and O-RAN integration.
\Cref{sec:wms3m} presents the WM--MS$^{3}$M architecture and the leakage-safe training/inference pipeline.
\Cref{sec:perf-numerical} reports quantitative results, uncertainty diagnostics, what-if forecasting, and MPC/CEM planning.
We conclude in \Cref{sec:conclusion}.

\begin{figure*}[t]
\centering
\includegraphics[width=\linewidth]{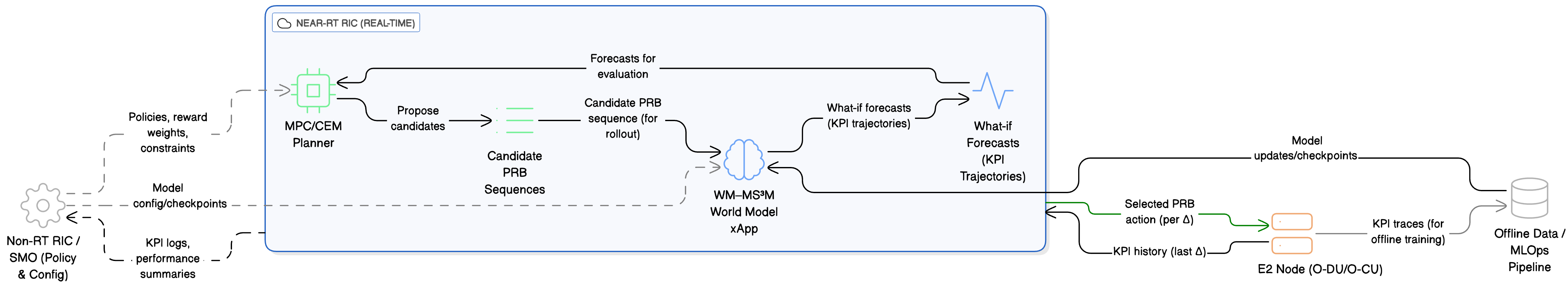}
\caption{Agentic world-modeling pipeline in the O-RAN Near-RT RIC. Aggregated KPIs and PRB actions from the E2 node feed the WM--MS$^{3}$M world model, which (i) provides calibrated one-step and short-horizon forecasts for factual or counterfactual PRB sequences, and (ii) serves as the dynamics backbone for an MPC/CEM planner that evaluates candidate PRB trajectories and outputs the next Near-RT control action.}

\label{fig:system-modeling}
\end{figure*}

\section{System Model}
\label{sec:system-model}

We adopt the O\mbox{-}RAN architecture in Figure~\ref{fig:system-modeling}, where an \emph{agentic} controller is realized as the composition of a learned \emph{world model} xApp and an outer-loop MPC/CEM planner. Both run in the Near-RT RIC and interact with the E2 node \cite{oran_architecture_overview, etsi_ts_104040}, the Non-RT RIC/service management and orchestration (SMO), and an offline data/machine learning operations (MLOps) pipeline. In this setting, the E2 node (i.e., O-RAN distributed unit (O-DU) and O-RAN central unit (O-CU)) executes scheduling and lower-layer procedures. Every aggregation period $\Delta$ (typically on the order of $10$--$1000$\,ms), it sends aggregated KPI history for the last interval (KPI history (last~$\Delta$) in Figure~\ref{fig:system-modeling}) to the Near-RT RIC over the E2 interface. It applies the selected PRB action produced by the planner (Selected PRB action (per~$\Delta$)). At the same time, the E2 node produces KPI traces that are exported to the offline data/MLOps pipeline (KPI traces (for offline training)) and later used to retrain or fine-tune the world model. In a live deployment, these KPIs would be delivered as aggregated telemetry over the E2 interface from the O-DU and O-CU. In our experiments, we instead use an \emph{offline dataset} of such KPIs collected from an O\mbox{-}RAN test setup, as described in \Cref{subsec:Data-Collection}.

The Near-RT RIC hosts both the WM--MS$^{3}$M world model xApp and the MPC/CEM planner. At each decision time $t$, the Near-RT RIC receives a window of aggregated KPIs and PRB usage from the E2 node and assembles them into a causal input window. The planner proposes candidate PRB sequences (the Candidate PRB sequences block in Figure~\ref{fig:system-modeling}) and sends them, together with the KPI window, to the world model. The world model evaluates each candidate by rolling out short-horizon, action-conditioned KPI trajectories and returning ``Forecasts for evaluation'' and ``What-if forecasts (KPI trajectories)''. The planner then scores these trajectories according to an operator-specified reward and selects the next PRB action, which is sent back to the E2 node as the ``Selected PRB action (per~$\Delta$)''.

The Non-RT RIC/SMO supervises this Near-RT loop at a slower timescale. It configures high-level policies, reward weights, and constraints that define the Near-RT objective for the planner, as indicated by the arrow ``Policies, reward weights, constraints'' feeding into the MPC/CEM block in Figure~\ref{fig:system-modeling}. It also distributes model configurations and checkpoints to the world model xApp (``Model config/checkpoints'' and ``Model updates/checkpoints''), typically after offline retraining on the KPI traces produced by the E2 node and processed in the offline data/MLOps pipeline. In the other direction, the Near-RT RIC reports KPI logs and performance summaries to the Non-RT RIC/SMO for monitoring and longer-term optimization.

Let $\bx_t\in\R^{F}$ denote the KPI vector aggregated over $[(t-1)\Delta,t\Delta)$ for a given sector (see Table~\ref{tab:notation-wmms3m} for major notations). It includes radio-quality and traffic/load indicators (e.g., RSRP, SINR, CQI, SE, BLER, delay, PRB usage). The Near-RT controller acts through a scalar control $u_t \equiv \mathrm{PRB}_t \in \R$, representing the scheduled PRBs in period $t$. We model the radio environment as a controlled, partially observed stochastic process. A latent state $x_t$ summarizes unobserved propagation, interference, and traffic conditions. The dynamics follow a generic state-space form
\begin{equation}
\label{eq:sys-ssm}
x_{t+1} = f_\theta(x_t,u_t) + w_t,\qquad
\bx_t = h_\theta(x_t) + v_t,
\end{equation}
where $w_t$ and $v_t$ capture stochastic variability and unmodeled effects. We do \emph{not} attempt to reconstruct $x_t$ explicitly. Instead, all predictions and planning are based on a causal summary of the observable history, which is exactly what is available to an xApp at the Near-RT RIC. Let $\cF_t$ denote the $\sigma$-algebra generated by $\{\bx_\tau,u_\tau\}_{\tau\le t}$. At time $t$, the Near-RT RIC has access to a standardized lookback window of length $L$,

\begingroup
\footnotesize
\begin{equation}
    \X_t = (\bx_{t-L+1},\dots,\bx_t)\in\R^{L\times F},\qquad
u_{t-L+1:t}\in\R^{L\times 1},
\end{equation}
\endgroup
and the one-step-ahead target $\by_{t+1}\in\R^{O},\quad O\in\{1, F\}$, where $\by_{t+1}$ may represent a single KPI (e.g., RSRP) or the full KPI vector. By construction, $(\X_t,u_{t-L+1:t})$ is $\cF_t$-measurable and $\by_{t+1}$ is $\cF_{t+1}$-measurable. Any predictor of the form $f_\theta:\ (\X_t,u_{t-L+1:t}) \mapsto \widehat{\by}_{t+1}$ is thus strictly causal and deployable in the Near-RT control loop. To keep planning within the region supported by the data and by scheduler limits, we restrict actions to a data-driven admissible set $\mathcal{U}\subset\R$. On the \emph{training} split, we compute the empirical $5$th and $95$th percentiles of the PRB distribution and define $[a_{\mathrm{lo}},a_{\mathrm{hi}}] \subset \R,~\mathcal{U} = [a_{\mathrm{lo}},a_{\mathrm{hi}}]$.

All candidate actions during planning are sampled and clamped in the \emph{standardized} space using the same affine scaling as for training; only the final selected action is mapped back to physical PRB units before being issued toward the scheduler. Over a planning horizon $H$, a candidate control path $\mathbf{a} = (u_t,u_{t+1},\dots,u_{t+H-1}) \in \mathcal{U}^{H}$, induces a distribution over future KPIs, $p_\theta\!\big(\by_{t+h}(\mathbf{a}) \,\big|\, \X_t,u_{t-L+1:t}\big)$, where $\by_{t+h}(\mathbf{a})$ denotes the random KPI vector at horizon $t{+}h$ under the hypothetical PRB path~$\mathbf{a}$, as represented by the learned world model. Within the Near-RT RIC, the WM--MS$^{3}$M xApp exposes two core operations to other xApps, rApps, or to the SMO, (i) \textbf{Factual one-step prediction:} $p_\theta(\by_{t+1}\mid \X_t,u_{t-L+1:t})$, returning calibrated means, variances, and optionally quantiles for the next KPI(s) given the factual control history and (ii) \textbf{Counterfactual (what--if) prediction:} for any hypothetical future control path $\mathbf{a}\in\mathcal{U}^H$,

\begingroup
\footnotesize
\begin{equation}
    \big\{p_\theta(\by_{t+h}(\mathbf{a})\mid \X_t,u_{t-L+1:t})\big\}_{h=1}^{H},
\end{equation}
\endgroup
providing action-conditioned short-horizon forecasts for planning and offline policy screening. An outer-loop planner (e.g., MPC/CEM) utilizes these predictive distributions, scores candidate $\mathbf{a}$ according to an operator-specified reward (trading SINR/SE/RSRP against BLER/Delay and PRB cost), and outputs only the first control $u_t^\star$ in a receding-horizon fashion, consistent with Near-RT RIC execution.

\begin{mdframed}[
  backgroundcolor=gray!5,
  linecolor=gray!50,
  linewidth=0.5pt,
  roundcorner=2pt,
  innerleftmargin=8pt,
  innerrightmargin=8pt,
  innertopmargin=6pt,
  innerbottommargin=6pt,
  skipabove=6pt,
  skipbelow=6pt,
  userdefinedwidth=\columnwidth,
  align=center
]
\centering
\footnotesize
\(
  \underbrace{\text{Predict / Imagine}}_{\text{WM--MS$^{3}$M (world model) xApp}}
  \;\longrightarrow\;
  \underbrace{\text{Choose}}_{\text{planner / control policy}} =\; \; \;\text{Agentic Control}
\)\\[3pt]
\scriptsize This separation of prediction/imagination and \emph{choice} is precisely what we mean by an \emph{agentic world model} for 6G O\mbox{-}RAN.

\end{mdframed}

\begin{table}[p]
\centering
\scriptsize
\setlength{\tabcolsep}{3pt}
\caption{Major Notations.}
\label{tab:notation-wmms3m}
\resizebox{\columnwidth}{!}{%
\begin{tabular}{lll}
\toprule
\textbf{Symbol} & \textbf{Description} & \textbf{Size/Type} \\
\midrule
\multicolumn{3}{l}{\emph{\textbf{Sets, probability, and operators}}}\\
$(\Omega,\cF,\PP)$ & Probability space & --- \\
$\E[\cdot],\ \Var[\cdot],\ \Cov[\cdot]$ & Expectation, variance, covariance & scalar/matrix \\
$\KL(\cdot\|\cdot)$ & Kullback--Leibler divergence & scalar \\
$\N(\mu,\Sigma)$ & (Multivariate) normal distribution & distribution \\
$\R,\ \mathbb{Z}_+$ & Reals; nonnegative integers & sets \\
$\Tr(\cdot)$ & Trace of a matrix & scalar \\
$\bigO(\cdot)$ & Big-\!O growth notation & --- \\
$\cF_t$ & Information $\sigma$-algebra up to time $t$ & $\sigma$-algebra \\
$\diag(\cdot)$ & Diagonal matrix from a vector & matrix \\
$\norm{\cdot},\ \abs{\cdot},\ \ip{\cdot}{\cdot}$ & Norm, absolute value, inner product & --- \\
$\odot$ & Hadamard (elementwise) product & --- \\
\midrule
\multicolumn{3}{l}{\emph{\textbf{Time, indices, and dimensions}}}\\
$F$ & Number of KPIs (features) & $\mathbb{Z}_+$ \\
$H$ & Planning/rollout horizon (future steps) & $\mathbb{Z}_+$ \\
$L$ & Lookback window (past steps) & $\mathbb{Z}_+$ \\
$L_\ell$ & Number of MS$^{3}$M layers & $\mathbb{Z}_+$ \\
$L_k$ & Kernel support length (taps) & $\mathbb{Z}_+$ \\
$M$ & SSM mixture components (time scales) & $\mathbb{Z}_+$ \\
$N$ & Per-channel SSM state size & $\mathbb{Z}_+$ \\
$O$ & Output size ($1$ univariate; $F$ multivariate) & $\mathbb{Z}_+$ \\
$S$ & MC samples for predictive averaging & $\mathbb{Z}_+$ \\
$T$ & Series length (timestamps) & $\mathbb{Z}_+$ \\
$\mathcal I_{\rm tr},\mathcal I_{\rm va},\mathcal I_{\rm te}$ & Train/val/test index sets & subsets of $\{1{:}T\}$ \\
$d$ & Embedding width (channels) & $\mathbb{Z}_+$ \\
$d_z$ & Latent world state dimension & $\mathbb{Z}_+$ \\
$h=\alpha d$ & GLU hidden width ($\alpha<2$) & $\mathbb{Z}_+$ \\
$i^\star$ & Target KPI index (when $O{=}1$) & index \\
$i_a$ & Action (PRB) feature index & index \\
\midrule
\multicolumn{3}{l}{\emph{\textbf{Data, windows, and scaling}}}\\
$\X_t=(\bx_{t-L+1},\dots,\bx_t)$ & Input window & $\R^{L\times F}$ \\
$\bm\mu_x,\bm\sigma_x$ & Per-feature input scalers (train-only) & $\R^{F}$ \\
$\bx_t\in\R^{F}$ & KPI vector at time $t$ & vector \\
$\by_{t+1}$ & Next-step target & $\R^{O}$ \\
$\mathbf{A}$, $\widetilde{\mathbf{A}}$ & PRB history (orig/standardized) & $\R^{L\times 1}$ \\
$\mu_y,\sigma_y$ / $(\bm\mu_y,\bm\sigma_y)$ & Target scaler(s) (train-only) & scalar(s)/$\R^{F}$ \\
$\widetilde{\X},\widetilde{\by}$ & Standardized inputs/targets & as $\X,\by$ \\
$a_{\text{lo}},a_{\text{hi}}$ & PRB bounds (5th/95th pct., orig units) & scalars \\
\midrule
\multicolumn{3}{l}{\emph{\textbf{MS$^{3}$M embeddings, gating, and mixers}}}\\
$H^{(0)}$ & Embedded sequence (inputs $+$ PRB channel) & $\R^{L\times d}$ \\
$H^{(\ell)}$ & Layer-$\ell$ output & $\R^{L\times d}$ \\
$U^{(\ell)}$ & Depthwise causal SSM conv output (layer $\ell$) & $\R^{L\times d}$ \\
$W_{\mathrm{in}}$ & Input projection (features$+1\!\to\!d$) & $\R^{(F{+}1)\times d}$ \\
$W_{\uparrow,a}^{(\ell)},\,W_{\uparrow,g}^{(\ell)}$ & GLU up-projections (act/gate) & $\R^{d\times h}$ \\
$W_{\downarrow}^{(\ell)}$ & GLU down-projection & $\R^{h\times d}$ \\
$s^{(\ell)},\,g^{(\ell)}$ & squeeze--excitation squeeze vector and gate & $\R^{d},\ \R^{d}$ \\
$Y^{(\ell)}$ & Residual $+$ LayerNorm output & $\R^{L\times d}$ \\
$Z^{(\ell)}$ & GLU mixer output & $\R^{L\times d}$ \\
$\mathrm{LN}(\cdot)$ & Layer normalization & map $\R^{d}\to\R^{d}$ \\
$\varphi_{\text{nl}}(\cdot)$ & Pointwise nonlinearity (e.g., GELU/ReLU) & scalar-wise map \\
$\sigma(\cdot)$ & Sigmoid (gate nonlinearity) & scalar-wise map \\
$\bd$ & Last-step deterministic summary & $\R^{d}$ \\
\midrule
\multicolumn{3}{l}{\emph{\textbf{HiPPO--LegS kernels and discretization}}}\\
$A(\Delta t)$ & Bilinear (Tustin) discretization of $A_{\ct}$ & $\R^{N\times N}$ \\
$A_{\ct}$ & HiPPO--LegS continuous-time operator & $\R^{N\times N}$ \\
$B,C,D$ & Discrete SSM parameters (depthwise, per channel) & $\R^{d\times N},\R^{d\times N},\R^{d}$ \\
$\Delta t^{(\ell,m)}$ & Learned positive time step (per layer, component) & $\R_{>0}$ \\
$k^{(\ell)}_{c}[\cdot]$ & Mixture-summed taps for channel $c$ & $\R^{L_k}$ \\
$k^{(\ell,m)}_{c}[\tau]$ & Tap for channel $c$, component $m$, lag $\tau$ & scalar \\
\midrule
\multicolumn{3}{l}{\emph{\textbf{Latent world model and decoders}}}\\
$(\bmu_p,\bsigma_p^2)$ & Prior parameters $p(\bz|\bd)$ & $\R^{d_z},\R^{d_z}$ \\
$(\bmu_q,\bsigma_q^2)$ & Posterior parameters $q(\bz|\bd,\widetilde{\by}_{\text{full}})$ & $\R^{d_z},\R^{d_z}$ \\
$\Dec_{\text{full}}$ & Full next-step decoder & $\R^{d{+}d_z}\!\to\!\R^{F}$ \\
$\Dec_{\text{t}}$ & Target decoder (mean[+logvar]) & $\R^{d{+}d_z}\!\to\!\R^{O}$ (or $2O$) \\
$\bb=f_{\mathrm{ar}}(\widetilde{\bx}_t)$ & AR skip from last input (std.\ space) & $\R^{O}$ \\
$\bmu_t,\ \log\bsigma_t^2$ & Target-head mean, log-variance & $\R^{O},\ \R^{O}$ \\
$\bz$ & Stochastic latent world state & $\R^{d_z}$ \\
$\kappa,\ \alpha=\tanh(\kappa)$ & Skip gain (raw/bounded) & scalars \\
$\widehat{\by}_{\text{full}}$ & Full-head next-step (std.\ space) & $\R^{F}$ \\
$\widehat{\by}_{\text{t}}$ & Target-head mean with AR skip & $\R^{O}$ \\
\midrule
\multicolumn{3}{l}{\emph{\textbf{Training objectives and schedules}}}\\
$E_{\max},\ p,\ \texttt{tol}$ & Max epochs; patience; validation tolerance & integers/scalar \\
$\alpha_{\cdot}$ & Loss weights (recon/t/Huber/cons/roll) & scalars \\
$\beta_e$ & KL weight (annealed over epochs) & scalar \\
$\mathcal{L}_{\mathrm{cons}}$ & Decoder-consistency penalty (target coord.) & scalar \\
$\mathcal{L}_{\mathrm{hub}}$ & Huber auxiliary penalty & scalar \\
$\mathcal{L}_{\mathrm{kl}}$ & KL($q\|\!p$) latent regularizer & scalar \\
$\mathcal{L}_{\mathrm{recon}}$ & Full-head MSE (fixed-variance Gaussian) & scalar \\
$\mathcal{L}_{\mathrm{roll}}$ & (Optional) aligned rollout penalty & scalar \\
$\mathcal{L}_{\mathrm{t}}$ & Target-head NLL (hetero.) or MSE (homo.) & scalar \\
$\pi_e$ & Posterior mixing prob.\ (scheduled) & scalar \\
$c_{\max},\ \lambda$ & Grad clip threshold; weight decay & scalars \\
$p_{\mathrm{cd}},\ \sigma_{\mathrm{in}}$ & Feature-channel dropout; input noise & scalars \\
\midrule
\multicolumn{3}{l}{\emph{\textbf{Inference, uncertainty, and de-standardization}}}\\
$\overline{\by}_{\text{t}}$ & MC-averaged target prediction (std.) & $\R^{O}$ \\
$\widehat{\Var}[\widetilde{\by}^{(O)}]$ & Predictive variance (aleatoric+epistemic) & $\R^{O}$ \\
$\widehat{\by}$ & De-standardized prediction (phys.\ units) & $\R^{O}$ \\
\midrule
\multicolumn{3}{l}{\emph{\textbf{Counterfactuals and PRB-constrained planning (MPC/CEM)}}}\\
$P,\ \rho,\ I,\ K$ & Pop.\ size; elite frac.; iters; \# elites & integers/scalar \\
$\bm\mu,\bm\sigma$ & CEM mean/std over action sequences (std.\ space) & $\R^{H\times1}$ \\
$\lambda_{\mathrm{sm}}$ & Action-smoothness penalty weight & scalar \\
$\mathbf{R}$ & Cumulative reward over horizon & scalar \\
$\mathbf{w}$ & Reward weights [SINR, SE, BLER, Delay, PRB, RSRP] & $\R^{6}$ \\
$\tilde a_{\text{lo}},\tilde a_{\text{hi}}$ & PRB bounds in standardized space & scalars \\
$a_t^{\star}$ & Selected next PRB (orig units) & scalar \\
$u_t\equiv \text{PRB}_t$ & Action (PRBs) at step $t$ & scalar (or small vector) \\
\bottomrule
\end{tabular}%
}
\end{table}

\section{World-Modeled Multi-Scale Structured State-Space Mixture}
\label{sec:wms3m}

We introduce \emph{WM--MS$^{3}$M}, a strictly causal forecaster that augments multi-scale structured state-space mixtures with a small latent world model. Each layer applies depthwise causal filters derived from HiPPO--LegS \cite{gu2020hippo} dynamics at multiple time scales and then mixes channels via squeeze--excitation \cite{hu2018squeeze} gating and a gated linear unit (GLU) \cite{dauphin2017language}. A compact latent $\bz$ is inferred from history (prior) or history-plus-future (posterior, training-only) and conditions two decoders: a full next-step mean squared error (MSE) reconstruction head (fixed-variance surrogate) and a \emph{heteroscedastic} target head with an autoregressive (AR) \cite{dalal2019autoregressive} skip. Training is \emph{leakage-safe} (train-only scalers, chronological split) and uses KL annealing \cite{Cui2025GKL} and scheduled posterior mixing. Monte Carlo (MC) \cite{robert-casella-2004-mcsm} prior sampling averages target-head means at test time; uncertainty is available via a variance decomposition. Algorithms~\ref{alg:wms3m_train_compact_codealigned} and \ref{alg:wms3m_infer_compact_codealigned} summarize the procedures. For action-conditioned decision support, we perform short-horizon planning via MPC/CEM with deterministic prior-mean rollouts; see Algorithm~\ref{alg:wms3m_plan_cem}.

\subsection{Problem Setup and Causality}
\label{subsec:wms3m-setup}

Let $\{\bx_t\}_{t=1}^{T}$ be a multivariate KPI time series with $\bx_t \in \R^{F}$. 
Fix a lookback horizon $L$ and construct input–target pairs

\begingroup\footnotesize
\begin{equation}
\label{eq:wms3m-windowing}
\X_t = (\bx_{t-L+1},\ldots,\bx_t) \in \R^{L\times F}, 
\qquad 
\by_{t+1} \in \R^{O},
\end{equation}
\endgroup
where $O\in\{1,F\}$ denotes, respectively, a univariate target (e.g., one KPI) or a multivariate target (all $F$ KPIs) \cite{Benidis2022DeepTimeSeriesSurvey, Salinas2020DeepAR}.

On the \emph{training} split, we fit per-feature standardization statistics and reuse them across all splits:

\begingroup\footnotesize
\begin{equation}
\widetilde{\X} = (\X - \bm\mu_x)\oslash \bm\sigma_x, 
\qquad
\widetilde{\by}=
\begin{cases}
(\by_{i^\star}-\mu_y)/\sigma_y, & O=1,\\[3pt]
(\by-\bm\mu_y)\oslash \bm\sigma_y, & O=F.
\end{cases}
\end{equation}
\endgroup

Since $\X_t$ is $\cF_t$-measurable (depends only on observations up to time $t$) and $\by_{t+1}$ lies one step ahead, any measurable predictor $f_\theta(\widetilde{\X}_t)$ is causal. 
\emph{(Alg.~\ref{alg:wms3m_train_compact_codealigned}, line 1; Alg.~\ref{alg:wms3m_infer_compact_codealigned}, line 1).}

\subsection{HiPPO--LegS Kernels, Bilinear Discretization, and Multi-Scale Mixture}
\label{subsec:wms3m-kernels}
\subsubsection{Continuous-time Template}
The HiPPO--LegS operator $A_{\ct}\in\R^{N\times N}$ and reference input $B_{\mathrm{ref}}\in\R^{N}$ are

\begin{subequations}
\label{eq:legs}
\begingroup
\footnotesize
\setlength{\abovedisplayskip}{6pt}
\setlength{\belowdisplayskip}{6pt}
\setlength{\jot}{2pt}
\begin{align}
(\Act)_{ij} &=
\begin{cases}
-\sqleg{i}{j}, & i>j,\\
-(i+1),        & i=j,\\
\ \,0,         & i<j,
\end{cases}\\[2pt]
(\Bref)_{i} &= \sqrt{\leg{i}}\,.
\end{align}
\endgroup
\end{subequations}

For a single channel with input $u(t)$ and state $\bs(t)\!\in\!\R^N$:
$\dot{\bs}(t)\!=\!A_{\ct}\bs(t)+B\,u(t),\ y(t)\!=\!C\bs(t)+D\,u(t)$, with learnable $(B,C,D)$ per channel.

\subsubsection{Bilinear (Tustin) Discretization}
For $\Delta t>0$,

\begingroup\footnotesize
\begin{equation}
\label{eq:tustin}
\begin{aligned}
A(\Delta t) &= \big(I - \tfrac{\Delta t}{2}A_{\ct}\big)^{-1}\big(I + \tfrac{\Delta t}{2}A_{\ct}\big),\\
k_c[0] &= C_c B_c + D_c,\qquad
k_c[\tau] = C_c\,A(\Delta t)^{\tau} B_c ,
\end{aligned}
\end{equation}
\endgroup
where $c\in\{1,\dots,d\}$ indexes channels. A matrix $A_{\ct}$ is called \emph{Hurwitz} if all its eigenvalues
have strictly negative real parts, which ensures asymptotic stability of the continuous-time LTI system
$\dot{x}=A_{\ct}x$~\cite{Hespanha2018LinearSystems}. A matrix $A$ is called \emph{Schur-stable} if all its
eigenvalues lie strictly inside the unit disk, i.e., $|\lambda_i(A)|<1$, which is the standard stability notion
for discrete-time systems $x_{k+1}=Ax_k$~\cite{Hespanha2018LinearSystems}. For the bilinear (Tustin)
discretization in~\eqref{eq:tustin}, standard results on Padé/bilinear transforms imply that if $A_{\ct}$ is
Hurwitz, then $A(\Delta t)$ is Schur-stable for any $\Delta t>0$~\cite{Sajja2010BilinearLyapunov,Shorten2011PadeQuadratic},
so the tap energy decays geometrically and a finite kernel length $L_k$ is a controlled truncation.

\subsubsection{Depthwise Causal Convolution and Mixture}
Let $\widetilde{\X}\mapsto H^{(0)}=\widetilde{\X}W_{\mathrm{in}}\in\R^{L\times d}$. At layer $\ell$, we form $M$ components with learned $\Delta t^{(\ell,m)}\!>\!0$ and taps $k^{(\ell,m)}_c[\cdot]\in\R^{L_k}$ and then sum them

\begingroup\footnotesize
\begin{equation}
    k^{(\ell)}_c[\cdot]=\sum_{m=1}^{M}k^{(\ell,m)}_c[\cdot],\qquad
    U^{(\ell)}_{t,c}=\sum_{\tau=0}^{L_k-1}k^{(\ell)}_{c}[\tau]\,H^{(\ell-1)}_{t-\tau,c},
\end{equation}
\endgroup
where $c\in\{1,\dots,d\}$ indexes channels and we define $H^{(\ell-1)}_{s,c}=0$ for all $s<1$. In other words, we pad the sequence with $(L_k-1)$ zeros on the \emph{left} (pre-padding) and no padding on the right, so that $U^{(\ell)}_{t,c}$ depends only on $\{H^{(\ell-1)}_{s,c} : 1\le s\le t\}$ and never on future positions. We set $L_k{=}L$ in code and implement this as a depthwise 1D convolution with \texttt{groups}$=d$ and left zero-padding, matching the theoretical per-channel SSM. \emph{(Alg.~\ref{alg:wms3m_train_compact_codealigned}, line 3 and lines 4–6).}

\subsubsection{Positivity Map for \texorpdfstring{$\Delta t$}{Delta t}}
We parameterize $\Delta t$ as $\Delta t=\phi(\tau)$ with

\begingroup\footnotesize
\begin{equation}
    \phi(\tau)=\log\big(1+\exp(\tau)\big)+\epsilon,\qquad \epsilon>0,
\end{equation}
\endgroup
so each learned time scale is strictly positive. The function $\phi(\tau)$ is the standard
\emph{softplus} nonlinearity (a smooth approximation of $\max\{0,\tau\}$), which maps
$\mathbb{R}\to(0,\infty)$ and has well-behaved derivatives for both negative and positive
inputs. This choice ensures that all $\Delta t$ used in the bilinear (Tustin) discretization
remain strictly positive (via the small offset $\epsilon$), avoiding invalid or numerically
degenerate step sizes while allowing unconstrained optimization in the parameter $\tau$.

\subsection{Channel Gating, GLU Mixing, and Normalization}
\label{subsec:wms3m-mix}
We compute a global squeeze--excitation gate $g^{(\ell)}\!\in\!(0,1)^d$ from $s^{(\ell)}=\frac{1}{L}\sum_t H^{(\ell-1)}_{t,\cdot}$ and modulate $\widehat{H}^{(\ell)}=U^{(\ell)}\odot g^{(\ell)}$. Residual and per-time-step layer norm yield $Y^{(\ell)}=\mathrm{LN}(H^{(\ell-1)}+\widehat{H}^{(\ell)})$. A compact GLU mixer with hidden $h=\alpha d$ produces

\begingroup \footnotesize\begin{equation}
\begin{aligned}
Z^{(\ell)} &=
W_{\downarrow}^{(\ell)}
\Big(
  \phi\!\big(W_{\uparrow,a}^{(\ell)} Y^{(\ell)}\big)
  \odot
  \sigma\!\big(W_{\uparrow,g}^{(\ell)} Y^{(\ell)}\big)
\Big),\\
H^{(\ell)} &= \mathrm{LN}\!\big(Y^{(\ell)} + Z^{(\ell)}\big).
\end{aligned}
\end{equation}\endgroup
After $L_\ell$ layers, we take the \emph{last} time embedding $\bd=H^{(L_\ell)}_{L,\cdot}\in\R^d$.
\emph{(Alg.~\ref{alg:wms3m_train_compact_codealigned}, line 10).}


\begin{algorithm}[t!]
\scriptsize
\caption{WM--MS\textsuperscript{3}M — Leakage-Safe Training}
\label{alg:wms3m_train_compact_codealigned}
\begin{algorithmic}[1]
\Require $\{\mathbf{x}_t\}_{t=1}^T$, $\mathbf{x}_t\!\in\!\R^{F}$; window $L$; output $O\!\in\!\{1,F\}$; target index $i^\star$; \textbf{action index $i_a$ (PRBs)}; state $N$; mixture $M$; layers $L_{\ell}$; taps $L_k{:=}L$; width $d$; latent $d_z$; HiPPO--LegS $A_{\ct}\!\in\!\R^{N\times N}$; splits $\mathcal I_{\mathrm{tr}},\mathcal I_{\mathrm{va}}$; $E_{\max}$; patience $p$; base LR steps $\{\gamma_e\}$; weight decay $\lambda$; grad clip $c_{\max}$; KL schedule $(\beta_{\mathrm{start}},\beta_{\mathrm{end}},E_{\mathrm{kl}})$; posterior mix $(\pi_0,\pi_1)$; augs: channel-drop $p_{\mathrm{cd}}$, input-noise $\sigma_{\mathrm{in}}$
\Ensure $\theta^\star$; scalers $(\bm\mu_x,\bm\sigma_x)$, $(\mu_y,\sigma_y)$, $(\bm\mu_y,\bm\sigma_y)$; action bounds $[a_{\text{lo}},a_{\text{hi}}]$
\State \textbf{Data \& scalers (train-only fit).} Build $(\mathbf{X}_t,\mathbf{y}_{t+1})$ with $\mathbf{X}_t\!\in\!\R^{L\times F}$, $\mathbf{y}_{t+1}\!\in\!\R^F$. On $\mathcal I_{\mathrm{tr}}$, compute $\bm\mu_x,\bm\sigma_x$; if $O{=}1$ compute $(\mu_y,\sigma_y)$ else $(\bm\mu_y,\bm\sigma_y)$. Scale: $\widetilde{\mathbf{X}}=(\mathbf{X}-\bm\mu_x)\oslash\bm\sigma_x$, and $\widetilde{\mathbf{y}}$ accordingly. \textit{Extract scaled PRB history} $\widetilde{\mathbf{A}}\in\R^{L\times 1}$ as column $i_a$ of $\widetilde{\mathbf{X}}$. \textit{Record PRB action bounds in original units} $[a_{\text{lo}},a_{\text{hi}}] \gets [\text{pct}_5, \text{pct}_{95}]$ of train PRBs.
\State \textbf{Params.} $W_{\mathrm{in}}\!\in\!\R^{(F{+}1)\times d}$ projects concatenated $[\widetilde{\mathbf{X}},\widetilde{\mathbf{A}}]$; MS\textsuperscript{3}M blocks $\{\mathcal{S}^{(\ell)}\}_{\ell=1}^{L_\ell}$. For each $(\ell,m)$: learn row-wise $B^{(\ell,m)},C^{(\ell,m)}\!\in\!\R^{d\times N}$, $D^{(\ell,m)}\!\in\!\R^{d}$, and $\Delta t^{(\ell,m)}=\phi(\tau^{(\ell,m)})$ (softplus). Share $A_{\ct}$. Squeeze--Excitation gates, GLU mixers, LayerNorms. Prior/posterior MLPs $\to(\bm\mu,\log\bm\sigma^2)\!\in\!\R^{d_z}$. Decoders: $\mathrm{Dec}_{\text{full}}\!:\R^{d{+}d_z}\!\to\!\R^F$, $\mathrm{Dec}_{\text{t}}\!:\R^{d{+}d_z}\!\to\!\R^O$. AR skip $\mathbf{b}=f_{\mathrm{ar}}(\widetilde{\mathbf{X}}_{L,\cdot})$, gain $\kappa$.

\WMSection{WORLD MODELING (MS\textsuperscript{3}M dynamics + latent \(\mathbf{z}\))}
\State \textbf{SSM taps (row-wise per channel).} $A(\Delta t)=\big(I-\frac{\Delta t}{2}A_{\ct}\big)^{-1}\big(I+\frac{\Delta t}{2}A_{\ct}\big)$;\;
$k_c[0]=C_c B_c + D_c$, $k_c[\tau]=C_cA^{\tau}B_c$, $\tau{=}1{:}L_k{-}1$;\; mixture $k^{(\ell)}_c=\sum_{m=1}^{M}k^{(\ell,m)}_c$.
\State \textbf{Forward $f_\theta$ (window, \underline{action-conditioned}).} $H^{(0)}=[\widetilde{\mathbf{X}},\widetilde{\mathbf{A}}]W_{\mathrm{in}}$; (train only) add Gaussian input noise $\mathcal{N}(0,\sigma_{\mathrm{in}}^2)$ and feature-channel dropout with prob.\ $p_{\mathrm{cd}}$.
\For{$\ell=1{:}L_{\ell}$}
  \State $U^{(\ell)}_{t,c}=\sum_{\tau=0}^{L_k-1}k^{(\ell)}_{c}[\tau]\;H^{(\ell-1)}_{t-\tau,c}$ \Comment{Depthwise causal conv (left padding, groups$=d$)}
  \State $s^{(\ell)}=\frac{1}{L}\sum_{t}H^{(\ell-1)}_{t,\cdot}$;\; $g^{(\ell)}=\sigma(W_2^{(\ell)}\phi(W_1^{(\ell)}s^{(\ell)}))$ \Comment{squeeze--excitation gate}
  \State $Y^{(\ell)}=\mathrm{LN}(H^{(\ell-1)}+U^{(\ell)}\odot g^{(\ell)})$ \Comment{Residual+norm}
  \State $Z^{(\ell)}=W_{\downarrow}^{(\ell)}(\phi(W_{\uparrow,a}^{(\ell)}Y^{(\ell)})\odot\sigma(W_{\uparrow,g}^{(\ell)}Y^{(\ell)}))$;\; $H^{(\ell)}=\mathrm{LN}(Y^{(\ell)}+Z^{(\ell)})$ \Comment{GLU mix}
\EndFor
\State $\mathbf{d}=H^{(L_\ell)}_{L,\cdot}$;\; $(\bm\mu_p,\log\bm\sigma_p^2)=\mathrm{MLP}_p(\mathbf{d})$
\State With prob.\ $\pi_e$ (per mini-batch) use posterior $(\bm\mu_q,\log\bm\sigma_q^2)=\mathrm{MLP}_q([\mathbf{d};\widetilde{\mathbf{y}}_{\text{full}}])$ else use prior; sample $\mathbf{z}$ by reparam.
\State $\widehat{\mathbf{y}}_{\text{full}}=\mathrm{Dec}_{\text{full}}([\mathbf{d};\mathbf{z}])$;\; target: $(\bm\mu_t,\log\bm\sigma_t^2)=\mathrm{Dec}_{\text{t}}([\mathbf{d};\mathbf{z}])$ (or $\bm\mu_t$ if homoscedastic);\; $\widehat{\mathbf{y}}_{\text{t}}=\bm\mu_t+\tanh(\kappa)\mathbf{b}$

\WMSection{OBJECTIVES \& SCHEDULES}
\State \textbf{Loss (mini-batch $\mathcal B\subset\mathcal I_{\mathrm{tr}}$).} $\mathcal{L}_{\mathrm{recon}}=\frac{1}{|\mathcal B|}\sum_{t}\|\widehat{\mathbf{y}}_{\text{full},t}-\widetilde{\mathbf{y}}_{\text{full},t}\|_2^2$
\If{heteroscedastic}
  \State $\mathcal{L}_{\mathrm{t}}=\frac{1}{|\mathcal B|}\sum_{t}\tfrac{1}{2}\!\left(\log\bm\sigma_t^2+\frac{(\widetilde{\mathbf{y}}^{(O)}_{t}-\widehat{\mathbf{y}}_{\text{t},t})^2}{\exp(\log\bm\sigma_t^2)}+\log 2\pi\right)$ \Comment{clamp $\log\bm\sigma_t^2\in[-8,8]$}
\Else
  \State $\mathcal{L}_{\mathrm{t}}=\frac{1}{|\mathcal B|}\sum_{t}\|\widehat{\mathbf{y}}_{\text{t},t}-\widetilde{\mathbf{y}}^{(O)}_{t}\|_2^2$
\EndIf
\State $\mathcal{L}_{\mathrm{hub}}=\frac{1}{|\mathcal B|}\sum_{t}\mathrm{Huber}_\delta(\widehat{\mathbf{y}}_{\text{t},t}-\widetilde{\mathbf{y}}^{(O)}_{t})$;\;
$\mathcal{L}_{\mathrm{cons}}=\frac{1}{|\mathcal B|}\sum_{t}\|\widehat{\mathbf{y}}_{\text{t},t}-\widehat{\mathbf{y}}_{\text{full},t}[i^\star]\|_2^2$
\State $\mathcal{L}_{\mathrm{kl}}=\frac{1}{|\mathcal B|}\sum_{t}\mathrm{KL}(\mathcal{N}(\bm\mu_q,\bm\sigma_q^2)\|\mathcal{N}(\bm\mu_p,\bm\sigma_p^2))$ \textit{(set $q{:=}p$ when prior is used)}
\State (Optional rollout) $\mathcal{L}_{\mathrm{roll}}=\frac{1}{K}\sum_{k=1}^K\|\widehat{\mathbf{y}}^{(k)}_{\text{full}}-\widetilde{\mathbf{y}}_{\text{full}}\|_2^2$
\State Schedules: $\beta_e=\beta_{\mathrm{start}}+(\beta_{\mathrm{end}}-\beta_{\mathrm{start}})\min(1,\tfrac{e}{E_{\mathrm{kl}}})$;\; $\pi_e=\pi_0+(\pi_1-\pi_0)\min(1,\tfrac{e}{E_{\mathrm{kl}}})$
\State Total: $\mathcal{L}=\alpha_{\mathrm{rec}}\mathcal{L}_{\mathrm{recon}}+\alpha_{\mathrm{t}}\mathcal{L}_{\mathrm{t}}+\alpha_{\mathrm{hub}}\mathcal{L}_{\mathrm{hub}}+\alpha_{\mathrm{cons}}\mathcal{L}_{\mathrm{cons}}+\beta_e\mathcal{L}_{\mathrm{kl}}+\alpha_{\mathrm{roll}}\mathcal{L}_{\mathrm{roll}}$

\State \textbf{Update \& early stop.} Mixed precision (CUDA); AdamW (weight decay $\lambda$); cosine warm restarts stepped \emph{per mini-batch} via fractional epoch; grad clip to $c_{\max}$. At epoch end compute $\mathcal{L}_{\mathrm{va}}$ (teacher-forced). If best improves by $>\texttt{tol}$: save $\theta^\star$ and reset counter; else increase. Stop when counter $\ge p$.
\State \textbf{Return} $\theta^\star$, scalers $(\bm\mu_x,\bm\sigma_x)$, $(\mu_y,\sigma_y)$, $(\bm\mu_y,\bm\sigma_y)$, and $[a_{\text{lo}},a_{\text{hi}}]$ \textit{(for Alg.~\ref{alg:wms3m_infer_compact_codealigned}/\ref{alg:wms3m_plan_cem})}.
\end{algorithmic}
\end{algorithm}

\subsection{Latent World Model and Dual Decoders}
\label{subsec:wms3m-world}
We parameterize a diagonal-Gaussian prior $p_\theta(\bz|\bd)=\mathcal{N}(\bmu_p,\diag(\bsigma_p^2))$ and a posterior
$q_\theta(\bz|\bd,\widetilde{\by}_{\text{full}})=\mathcal{N}(\bmu_q,\diag(\bsigma_q^2))$ via multilayer perceptrons (MLPs),

\begingroup \footnotesize\begin{equation}
    (\bmu_p,\log\bsigma_p^2)=\mathrm{MLP}_p(\bd),\qquad
(\bmu_q,\log\bsigma_q^2)=\mathrm{MLP}_q([\bd;\widetilde{\by}_{\text{full}}]).
\end{equation}\endgroup
Using the reparameterization trick, we form $[\bd;\bz]$ and decode

\begingroup \footnotesize\begin{equation}
\begin{aligned}
\widehat{\by}_{\text{full}} &= \Dec_{\text{full}}([\bd;\bz]) \in \R^{F},\\
(\bmu_t,\ \log\bsigma_t^{2}) &= \Dec_{\text{t}}([\bd;\bz]) \in \R^{O}\quad \text{(heteroscedastic)}.
\end{aligned}
\end{equation}\endgroup
A small learned function of the last standardized input (implemented as a two-layer MLP) serves as an AR-style skip with a bounded gain,

\begingroup \footnotesize\begin{equation}
    \bb=f_{\mathrm{ar}}(\widetilde{\bx}_t)\in\R^O,\qquad
\widehat{\by}_{\text{t}}=\bmu_t+\tanh(\kappa)\,\bb,
\end{equation}\endgroup
which improves short-horizon fidelity while preserving causality \emph{(Alg.~\ref{alg:wms3m_train_compact_codealigned}, lines 11-12)}. For numerical stability in the heteroscedastic head we clamp $\log\bsigma_t^{2}$ elementwise to $[-8,8]$ before computing the likelihood and gradients (see \S\ref{subsec:wms3m-world-math}).
\begin{algorithm}[t!]
\scriptsize
\caption{WM--MS\textsuperscript{3}M — Prior-Sampling Inference (PRB-Conditioned MC Averaging)}
\label{alg:wms3m_infer_compact_codealigned}
\begin{algorithmic}[1]
\Require From Alg.~\ref{alg:wms3m_train_compact_codealigned}: $\theta^\star$, scalers $(\bm\mu_x,\bm\sigma_x)$, $(\mu_y,\sigma_y)$ or $(\bm\mu_y,\bm\sigma_y)$; window $\mathbf{X}\!\in\!\R^{L\times F}$ (orig units); \textbf{extract PRB history} $\mathbf{A}=\mathbf{X}_{:,i_a}\in\R^{L\times 1}$; MC samples $S$; output size $O\!\in\!\{1,F\}$
\Ensure $\hat{\mathbf{y}}\!\in\!\R^{O}$ in original units
\State \textbf{Scale input.} $\widetilde{\mathbf{X}}=(\mathbf{X}-\bm\mu_x)\oslash\bm\sigma_x$;\; $\widetilde{\mathbf{A}}=(\mathbf{A}-\mu_{x,i_a})/\sigma_{x,i_a}$
\State \textbf{Deterministic summary.} $H^{(0)}=[\widetilde{\mathbf{X}},\widetilde{\mathbf{A}}]W_{\mathrm{in}}$; pass MS\textsuperscript{3}M blocks to get $\mathbf{d}=H^{(L_\ell)}_{L,\cdot}$
\State \textbf{Prior from $\theta^\star$.} $(\bm\mu_p,\log\bm\sigma_p^2)=\mathrm{MLP}_p(\mathbf{d})$
\For{$s=1{:}S$}
  \State Sample $\mathbf{z}^{(s)} \sim \mathcal{N}(\bm\mu_p,\mathrm{diag}(\bm\sigma_p^2))$
  \State Decode target mean $\bm\mu_t^{(s)}=\mathrm{Dec}_{\text{t}}([\mathbf{d};\mathbf{z}^{(s)}])$ \Comment{optionally use decoded log-variance for UQ}
  \State AR skip: $\mathbf{b}=f_{\mathrm{ar}}(\widetilde{\mathbf{X}}_{L,\cdot})$;\; $\widehat{\mathbf{y}}_{\text{t}}^{(s)}=\bm\mu_t^{(s)}+\tanh(\kappa)\mathbf{b}$
\EndFor
\State \textbf{MC mean (scaled).} $\overline{\mathbf{y}}_{\text{t}}=\tfrac{1}{S}\sum_{s=1}^{S}\widehat{\mathbf{y}}_{\text{t}}^{(s)}$
\If{$O{=}1$} \State \textbf{Invert scale.} $\hat{\mathbf{y}}=\mu_y+\sigma_y\,\overline{\mathbf{y}}_{\text{t}}$
\Else \State \textbf{Invert scale.} $\hat{\mathbf{y}}=\bm\mu_y+\bm\sigma_y\odot\overline{\mathbf{y}}_{\text{t}}$
\EndIf
\State \textbf{Return} $\hat{\mathbf{y}}$
\end{algorithmic}
\end{algorithm}
\subsection{Mathematical Details of the Latent World Model}
\label{subsec:wms3m-world-math}

\subsubsection{Variables and Conditionings}
At each time $t$, the standardized lookback window $\widetilde{\X}_t\!\in\!\R^{L\times F}$ is mapped by the SSM stack into a deterministic summary
$\bd_t \!=\! f_\theta(\widetilde{\X}_t)\in\R^{d}$ (the last-step embedding).
The (train-only) standardized next-step vector is $\widetilde{\by}_{\text{full},t}\!\in\!\R^{F}$,
and the (possibly univariate) standardized target is $\widetilde{\by}^{(O)}_t\!\in\!\R^{O}$ with $O\in\{1,F\}$.
A compact latent state $\bz_t\!\in\!\R^{d_z}$ captures information not determined by $\bd_t$ alone.

\subsubsection{Generative Parameterization (Conditional VAE View)}
We specify a diagonal-Gaussian prior and two conditional decoders

\begingroup \footnotesize\begin{equation}
p_\theta(\bz_t \mid \bd_t)
= \N\!\big(\bmu_p(\bd_t),\, \diag(\bsig_p^2(\bd_t))\big),
\label{eq:prior}
\end{equation}\endgroup
\begingroup \footnotesize\begin{equation}
p_\theta(\tilde{\by}_{\text{full},t} \mid \bd_t,\bz_t)
\propto
\exp\!\Big(-\tfrac{1}{2\sigma_{\mathrm{fixed}}^{2}}
\big\|\tilde{\by}_{\text{full},t}-\bmu_{\text{full}}([\bd_t;\bz_t])\big\|_2^2\Big),
\label{eq:full-like}
\end{equation}\endgroup
\begingroup \footnotesize\begin{equation}
\begin{aligned}
p_\theta(\tilde{\by}^{(O)}_t \mid \bd_t,\bz_t,\tilde{\bx}_{t})
&= \N\!\Big( \bmu_t([\bd_t;\bz_t]) + \alpha\,\bb_t,\; \diag(\bsig_t^2([\bd_t;\bz_t])) \Big),
\end{aligned}
\label{eq:target-like}
\end{equation}\endgroup
where $\bb_t=f_{\mathrm{ar}}(\widetilde{\bx}_{t})$ is a small MLP of the \emph{last} standardized input vector
and $\alpha=\tanh(\kappa)\in(-1,1)$ is a learned bounded gain.\footnote{When $O{=}1$, Equation \eqref{eq:target-like} uses a scalar variance $\sigma_t^2$; when $O{=}F$ we use a diagonal covariance.}
Equation~\eqref{eq:full-like} is implemented as MSE (i.e., a fixed-variance Gaussian surrogate) for the \emph{full} reconstruction head, and Equation \eqref{eq:target-like} for the \emph{target} head.

\subsubsection{Inference Model and Leakage Safety}
Training uses an amortized Gaussian posterior

\begingroup\footnotesize
\begin{equation}
q_\theta(\bz_t \mid \bd_t,\widetilde{\by}_{\text{full},t})
= \mathcal{N}\!\big(\bmu_q([\bd_t;\widetilde{\by}_{\text{full},t}]),\, \diag(\bsigma_q^2([\bd_t;\widetilde{\by}_{\text{full},t}]))\big),
\label{eq:posterior}
\end{equation}
\endgroup
in the standard variational autoencoder style with amortized inference
\cite{Kingma2014AutoEncoding}. The standardization parameters
$(\bm\mu_x,\bm\sigma_x)$ and $(\mu_y,\sigma_y)$ or $(\bm\mu_y,\bm\sigma_y)$ are
estimated on the \emph{training} split only and then kept fixed for
validation/test, to avoid look-ahead leakage in the time-series setting
\cite{Hyndman2018Forecasting}. For numerical stability in the
heteroscedastic target head, we clip $\log\bsigma_t^2$ to a bounded interval
(e.g., $[-8,8]$) before computing the likelihood, following common practice in
deep uncertainty-aware regression \cite{kendall2017uncertainties}.

\subsubsection{KL-annealed ELBO with Scheduled Posterior Mixing}
Introduce a Bernoulli switch $s_t\!\sim\!\mathrm{Bernoulli}(\pi_e)$ sampled \emph{per mini-batch}.
Define the chosen encoder

\begingroup \footnotesize\begin{equation}
    \tilde q_t(\bz_t)=
\begin{cases}
q_\theta(\bz_t \mid \bd_t,\widetilde{\by}_{\text{full},t}), & s_t=1\quad\text{(posterior phase)},\\
p_\theta(\bz_t \mid \bd_t), & s_t=0\quad\text{(prior phase)},
\end{cases}
\end{equation}\endgroup
so that, when $s_t\!=\!0$ (prior phase), we set $\tilde q_t \equiv p_\theta(\bz_t\mid\bd_t)$ and the KL term \emph{evaluates to zero} (KL between identical Gaussians).
With $\beta_e$ the KL weight at epoch $e$, the per-sample ELBO reads

\begingroup \footnotesize\begin{equation}
\begin{aligned}
\mathcal{L}_{\text{ELBO},t}
&=
\underbrace{\E_{\bz_t\sim \tilde q_t}\!\big[\log p_\theta(\widetilde{\by}_{\text{full},t}\mid \bd_t,\bz_t)\big]}_{\text{full reconstruction (MSE surrogate)}}
\\
&\quad+
\underbrace{\E_{\bz_t\sim \tilde q_t}\!\big[\log p_\theta(\widetilde{\by}^{(O)}_t \mid \bd_t,\bz_t,\widetilde{\bx}_{t})\big]}_{\substack{\text{target}\\\text{prediction}}}
\\
&\quad-
\beta_e\,
\underbrace{\KL\!\big(\tilde q_t(\bz_t)\;\|\;p_\theta(\bz_t\mid \bd_t)\big)}_{\text{latent regularization}}.
\end{aligned}
\label{eq:elbo}
\end{equation}\endgroup

The training loss in Equation~\eqref{eq:wms3m-total} is a weighted \emph{negative} ELBO surrogate:
$\mathcal{L}_{\mathrm{recon}}$ approximates the first term (fixed-variance Gaussian via MSE);
$\mathcal{L}_{\mathrm{t}}$ and $\mathcal{L}_{\mathrm{hub}}$ the second;
$\mathcal{L}_{\mathrm{kl}}$ the third; and $\mathcal{L}_{\mathrm{cons}}$ encourages agreement between the two decoders on the target coordinate.
Schedules $\beta_e$ and $\pi_e$ follow the linear ramps defined in \S\ref{subsec:wms3m-loss},

\begingroup \footnotesize\begin{equation}
\begin{split}
\KL\big(\cN_q\|\cN_p\big)
= \frac{1}{2}\sum_{j=1}^{d_z}\Bigg[
\frac{\sigma_{q,j}^2}{\sigma_{p,j}^2}
+ \frac{(\mu_{p,j}-\mu_{q,j})^2}{\sigma_{p,j}^2}
-1+ \log\frac{\sigma_{p,j}^2}{\sigma_{q,j}^2}
\Bigg].
\end{split}
\label{eq:diag-kl}
\end{equation}\endgroup

\subsubsection{Reparameterization and Gradients}
For either $\tilde q_t$, sample with the reparameterization technique:

\begingroup \footnotesize\begin{equation}
    \bz_t=\bmu_{\tilde q_t}+ \bsigma_{\tilde q_t}\odot \bepsilon,\qquad \bepsilon\sim\cN(\mathbf{0},I),
\end{equation}\endgroup
which yields unbiased, low-variance Monte Carlo estimates of the expectations in Equation \eqref{eq:elbo}.

\subsubsection{Predictive Distribution and MC Calibration}
At test time, the prior-only predictive for the standardized target is

\begingroup \footnotesize\begin{equation}
p_\theta(\widetilde{\by}^{(O)}_t \mid \bd_t,\widetilde{\bx}_t)
= \int p_\theta(\widetilde{\by}^{(O)}_t \mid \bd_t,\bz_t,\widetilde{\bx}_t)\;
p_\theta(\bz_t\mid \bd_t)\; d\bz_t.
\label{eq:prior-predictive}
\end{equation}\endgroup
We approximate Equation \eqref{eq:prior-predictive} by MC, draw $\bz_t^{(s)}\!\sim\!p_\theta(\bz_t\mid\bd_t)$, $s{=}1{:}S$, and set

\begingroup \footnotesize\begin{equation}
\begin{split}
\overline{\by}_{\text{t}}
=\frac{1}{S}\sum_{s=1}^{S}\Big(\bmu_t([\bd_t;\bz_t^{(s)}])+\alpha\,\bb_t\Big),\\
\widehat{\by}=
\begin{cases}
\mu_y+\sigma_y\,\overline{\by}_{\text{t}}, & O=1,\\
\bm\mu_y+\bm\sigma_y\odot \overline{\by}_{\text{t}}, & O=F.
\end{cases}
\end{split}
\end{equation}\endgroup

When the decoder in Equation \eqref{eq:target-like} is heteroscedastic,
an unbiased estimator of the \emph{standardized} predictive variance decomposes as

\begingroup \footnotesize\begin{equation}
\begin{aligned}
\widehat{\operatorname{Var}}\!\big[\widetilde{\boldsymbol{y}}^{(O)}_{t}\big]
&=
\underbrace{\tfrac{1}{S}\sum_{s=1}^{S}\boldsymbol{\sigma}_{t}^{2}\!\big([\boldsymbol{d}_{t};\boldsymbol{z}_{t}^{(s)}]\big)}_{\text{aleatoric}}\\
&\quad+
\underbrace{\tfrac{1}{S-1}\sum_{s=1}^{S}\!\left(\boldsymbol{\mu}_{t}\!\big([\boldsymbol{d}_{t};\boldsymbol{z}_{t}^{(s)}]\big)-\overline{\boldsymbol{\mu}}_{t}\right)^{2}}_{\text{epistemic (MC)}}\\
\overline{\boldsymbol{\mu}}_{t}
&=\tfrac{1}{S}\sum_{s=1}^{S}\boldsymbol{\mu}_{t}\!\big([\boldsymbol{d}_{t};\boldsymbol{z}_{t}^{(s)}]\big),
\end{aligned}
\label{eq:var-decomp}
\end{equation}\endgroup
with the AR skip $\alpha\bb_t$ deterministic given $\widetilde{\X}_t$.
Variance in original units follows by multiplying elementwise with $\sigma_y^2$ (or $\bm\sigma_y^2$).

\subsubsection{Consistency Between Decoders}
Let $i^\star$ denote the target index when $O{=}1$.
The consistency loss $\mathcal{L}_{\mathrm{cons}}$ penalizes the discrepancy between the \emph{target-head mean}
$\widehat{\by}_{\text{t},t}=\bmu_t+\alpha\,\bb_t$ and the corresponding coordinate of the full-head output,
i.e., $\big\|\widehat{\by}_{\text{t},t}-\widehat{\by}_{\text{full},t}[i^\star]\big\|_2^2$,
encouraging the two conditionals to agree on the target coordinate while allowing the full head to model
residual cross-feature structure useful for representation learning.

\subsubsection{AR Skip as a Small Learned Control}
Writing the target likelihood as
\(
\widetilde{\by}^{(O)}_t = \bmu_t([\bd_t;\bz_t]) + \alpha\,f_{\mathrm{ar}}(\widetilde{\bx}_{t}) + \bepsilon_t,\ \ \bepsilon_t\sim\cN(0,\diag(\bsigma_t^2)),
\)
the term $\alpha\,f_{\mathrm{ar}}(\widetilde{\bx}_{t})$ (with $\alpha=\tanh(\kappa)\!\in\!(-1,1)$) acts as a \emph{known exogenous control} computed from the most recent inputs by a small MLP, which improves short-horizon fidelity without violating causality. In practice, we clamp $\log\bsigma_t^2$ to $[-8,8]$ to avoid numerical pathologies.

The latent is identifiable up to affine transformations that can be absorbed by the decoders;
the prior/posterior are diagonal to avoid degenerate rotations.
Scale symmetries across $(\bd_t,\bz_t)$ are partially fixed by LayerNorm/squeeze--excitation and by the KL penalty Equation \eqref{eq:diag-kl}. Equations \eqref{eq:prior}–\eqref{eq:elbo} formalize WM--MS\textsuperscript{3}M as a \emph{causal, conditional VAE}
whose encoder switches between posterior and prior according to a scheduled Bernoulli, enabling stable training (via KL annealing)
and calibrated mean inference (via prior MC in Equations \eqref{eq:prior-predictive}–\eqref{eq:var-decomp}) without label leakage.

\subsection{Leakage-Safe Objectives and Schedules}
\label{subsec:wms3m-loss}
For a mini-batch $\mathcal{B}$, we combine five terms (all in standardized space):

\begingroup \footnotesize\begin{equation}
\mathcal{L}_{\mathrm{recon}}
=\frac{1}{|\mathcal{B}|}\sum_{t\in\mathcal{B}}\|\widehat{\by}_{\text{full},t}-\widetilde{\by}_{\text{full},t}\|_2^2,
\end{equation}\endgroup
\begingroup \footnotesize\begin{equation}
\mathcal{L}_{\mathrm{t}}
=\begin{cases}
\frac{1}{|\mathcal{B}|}\sum_t \tfrac{1}{2}\!\left(\log\bsigma_t^2+\frac{(\widetilde{\by}^{(O)}_t-\widehat{\by}_{\text{t},t})^2}{\exp(\log\bsigma_t^2)}+\log 2\pi\right), & \text{hetero},\\[3pt]
\frac{1}{|\mathcal{B}|}\sum_t \|\widehat{\by}_{\text{t},t}-\widetilde{\by}^{(O)}_t\|_2^2, & \text{homo},
\end{cases}
\end{equation}\endgroup
\begingroup \footnotesize\begin{equation}
\mathcal{L}_{\mathrm{hub}}
=\frac{1}{|\mathcal{B}|}\sum_t \mathrm{Huber}_\delta\!\left(\widehat{\by}_{\text{t},t}-\widetilde{\by}^{(O)}_t\right),
\end{equation}\endgroup
\begingroup \footnotesize\begin{equation}
\mathcal{L}_{\mathrm{cons}}
=\frac{1}{|\mathcal{B}|}\sum_t \left\|\widehat{\by}_{\text{t},t}-\widehat{\by}_{\text{full},t}[i^\star]\right\|_2^2,
\end{equation}\endgroup
\begingroup \footnotesize\begin{equation}
\mathcal{L}_{\mathrm{kl}}
=\frac{1}{|\mathcal{B}|}\sum_t \mathrm{KL}\!\left(\mathcal{N}(\bmu_q,\bsigma_q^2)\,\|\,\mathcal{N}(\bmu_p,\bsigma_p^2)\right),
\end{equation}\endgroup
where $\widetilde{\by}_{\text{full}}$ is the standardized next-step $F$-vector and $i^\star$ selects the target component when $O{=}1$. For the \emph{heteroscedastic} case lets the model predict a separate variance $\bsigma_t^2$ for each target (data-dependent noise level), while the \emph{homoscedastic} case corresponds to a standard fixed-variance Gaussian (or equivalently, an MSE loss) with constant noise across all samples \cite{kendall2017uncertainties}. The aligned rollout penalty averages teacher-forced multi-step reconstruction errors when aligned labels are provided:

\begingroup \footnotesize\begin{equation} 
\mathcal{L}_{\mathrm{roll}}=\frac{1}{K}\sum_{k=1}^K \left\|\widehat{\by}^{(k)}_{\text{full}}-\widetilde{\by}_{\text{full}}\right\|_2^2.
\end{equation}\endgroup

We use a linear schedule for the KL \emph{weight} and for the posterior-use
probability over training epochs $e$:

\begingroup\footnotesize
\begin{equation}
\begin{aligned}
\beta_e &=
\beta_{\mathrm{start}}
+ \big(\beta_{\mathrm{end}}-\beta_{\mathrm{start}}\big)\,
\min\!\left(1,\frac{e}{E_{\mathrm{kl}}}\right),\\
\pi_e &=
\pi_0
+ (\pi_1-\pi_0)\,
\min\!\left(1,\frac{e}{E_{\mathrm{kl}}}\right).
\end{aligned}
\end{equation}
\endgroup
Here $\beta_e$ is the epoch-dependent \emph{weight} applied to the KL regularizer
$\mathcal{L}_{\mathrm{kl}}$ in Equation \eqref{eq:wms3m-total}, and $\pi_e$ is the epoch-dependent
\emph{probability} of using the posterior encoder (instead of the prior) when
sampling the latent $\mathbf{z}_t$ during training. The overall objective is

\begingroup \footnotesize\begin{equation}
\label{eq:wms3m-total}
\begin{aligned}
\mathcal{L} &=
\alpha_{\mathrm{rec}}\,\mathcal{L}_{\mathrm{recon}}
+\alpha_{\mathrm{t}}\,\mathcal{L}_{\mathrm{t}}
+\alpha_{\mathrm{hub}}\,\mathcal{L}_{\mathrm{hub}}\\
&\quad
+\alpha_{\mathrm{cons}}\,\mathcal{L}_{\mathrm{cons}}
+\beta_e\,\mathcal{L}_{\mathrm{kl}}
+\alpha_{\mathrm{roll}}\,\mathcal{L}_{\mathrm{roll}}.
\end{aligned}
\end{equation}\endgroup
\subsubsection{Optimization and Early Stopping}
We use mixed-precision training, AdamW with weight decay $\lambda$, Cosine Warm Restarts, gradient-norm clipping $\|g\|\le c_{\max}$, and patience-based early stopping on validation loss. The cosine scheduler is stepped \emph{per mini-batch} using a fractional epoch counter (rather than once per epoch) to match the implementation. All scalers are fit on \emph{train only}; metrics (MSE, RMSE, and MAE) are reported after inverse standardization.
\emph{(Alg.~\ref{alg:wms3m_train_compact_codealigned}, lines 13–24; line 24 returns).}

\subsection{Prior-Sampling Inference (MC Averaging)}
\label{subsec:wms3m-infer}
At test time, compute $\bd$, form the prior $(\bmu_p,\bsigma_p^2)$ as in Equation \eqref{eq:prior}, draw $\bz^{(s)}\!\sim\!\mathcal{N}(\bmu_p,\diag(\bsigma_p^2))$, decode $\bmu_t^{(s)}$, add the AR skip, and average. The resulting predictor and scale inversion are:

\begingroup \footnotesize\begin{equation}
\begin{aligned}
\overline{\by}_{\mathrm{t}}
&= \frac{1}{S}\sum_{s=1}^{S}\!\left(\bmu_t([\bd;\bz^{(s)}])+\tanh(\kappa)\,\bb\right),\\
&\text{with }\;\bz^{(s)}\!\sim\!\mathcal{N}\!\left(\bmu_p(\bd),\diag(\bsigma_p^2(\bd))\right)
\\[4pt]
\widehat{\by}
&=
\begin{cases}
\mu_y+\sigma_y\,\overline{\by}_{\mathrm{t}}, & O=1,\\
\bm\mu_y+\bm\sigma_y\odot\overline{\by}_{\mathrm{t}}, & O=F.
\end{cases}
\end{aligned}
\label{eq:wms3m-mc-infer}
\end{equation}\endgroup

For uncertainty, use the variance decomposition in Equation \eqref{eq:var-decomp}.

\begin{algorithm}[t!]
\scriptsize
\caption{PRB Planning via MPC/CEM (Deterministic Prior-Mean Rollout)}
\label{alg:wms3m_plan_cem}
\begin{algorithmic}[1]
\Require From Alg.~\ref{alg:wms3m_train_compact_codealigned}: $\theta^\star$; \textbf{scalers} $(\bm\mu_x,\bm\sigma_x)$ for \emph{inputs/actions} and $(\bm\mu^{\text{full}}_y,\bm\sigma^{\text{full}}_y)$ for \emph{decoded frames}; action bounds $[a_{\text{lo}},a_{\text{hi}}]$ in original units; window $\mathbf{X}\!\in\!\R^{L\times F}$ (original units); action index $i_a$; horizon $H$; population $P$; elite fraction $\rho\!\in\!(0,1]$; iterations $I$; smoothness weight $\lambda_{\mathrm{sm}}$; \textbf{reward weights} $\mathbf{w}\in\R^{6}$ \emph{(learnable, initialized from config)} for \{SINR, SE, BLER, Delay, PRB, RSRP\} in scaled space.
\Ensure Next PRB $a_{t}^{\star}$ in original units
\State \textbf{Scale input \& compute scaled bounds.}
  \State $\widetilde{\mathbf{X}} \gets (\mathbf{X}-\bm\mu_x)\oslash \bm\sigma_x$;\; $\widetilde{\mathbf{A}} \gets \widetilde{\mathbf{X}}_{:,i_a:i_a{+}1}$
  \State $\tilde a_{\text{lo}} \gets \dfrac{a_{\text{lo}}-\mu_{x,i_a}}{\sigma_{x,i_a}},\quad \tilde a_{\text{hi}} \gets \dfrac{a_{\text{hi}}-\mu_{x,i_a}}{\sigma_{x,i_a}}$
  \State \textit{All CEM sampling, clamping, rewards, and rollouts are performed in standardized (scaled) space; the selected action is finally mapped back to original units.}
\State \textbf{Initialize CEM distribution.}
  \State $\bm\mu \in \R^{H\times 1}\gets \widetilde{A}_L\cdot \mathbf{1}$;\; $\bm\sigma \in \R^{H\times 1}\gets 0.5\cdot \mathbf{1}$
  \State $K \gets \max\!\big(1, \lfloor \rho P \rfloor\big)$
\Function{Rollout}{$\{\tilde{\mathbf{a}}^{(b)}\}_{b=1}^B$}
  \State Copy windows: $\widetilde{\mathbf{X}}^{(b)}\!\gets\!\widetilde{\mathbf{X}},\ \widetilde{\mathbf{A}}^{(b)}\!\gets\!\widetilde{\mathbf{A}}$;\; $\mathbf{R}^{(b)}\!\gets\!0$
  \For{$t=1{:}H$}
    \State \textbf{Deterministic latent (prior mean).} $\mathbf{d}^{(b)} \gets f_{\theta^\star}^{(d)}\!\big(\widetilde{\mathbf{X}}^{(b)},\widetilde{\mathbf{A}}^{(b)}\big)$;\;
           $(\bm\mu_p,\log\bm\sigma^2_p)\gets \mathrm{MLP}_p(\mathbf{d}^{(b)})$;\; $\mathbf{z}^{(b)}\gets \bm\mu_p$
    \State \textbf{Decode next frame (scaled).} $\widehat{\mathbf{y}}_{\text{full}}^{(b)} \gets \mathrm{Dec}_{\text{full}}\big([\mathbf{d}^{(b)};\mathbf{z}^{(b)}]\big)$
    \State \textbf{Apply candidate action.} $\tilde a_t^{(b)} \gets \tilde{\mathbf{a}}^{(b)}[t]$
    \State \textbf{Reward (scaled space).}
      \State Extract $\text{SINR},\text{SE},\text{BLER},\text{Delay},\text{RSRP}$ from $\widehat{\mathbf{y}}_{\text{full}}^{(b)}$ at their feature indices
      \State $\mathbf{f}_t^{(b)} \gets \big[\text{SINR},\text{SE},\text{BLER},\text{Delay},\tilde a_t^{(b)},\text{RSRP}\big]$
      \State $r_t^{(b)} \gets \langle \mathbf{w},\ \mathbf{f}_t^{(b)}\odot [\,1,\,1,\,-1,\,-1,\,-1,\,1\,] \rangle$
    \State \textbf{Smoothness penalty.} $s_t^{(b)} \gets \lambda_{\mathrm{sm}}\cdot \big|\tilde a_t^{(b)}-\tilde a_{t-1}^{(b)}\big|$ (with $\tilde a_0^{(b)}\!=\!\widetilde{A}^{(b)}_L$)
    \State \textbf{Accumulate \& advance window.} $\mathbf{R}^{(b)}\!\gets\!\mathbf{R}^{(b)}+r_t^{(b)}-s_t^{(b)}$
      \State $\widetilde{\mathbf{X}}^{(b)} \gets \text{concat}\big(\widetilde{\mathbf{X}}^{(b)}_{2:L,:},\ \widehat{\mathbf{y}}_{\text{full}}^{(b)}\big)$
      \State $\widetilde{\mathbf{A}}^{(b)} \gets \text{concat}\big(\widetilde{\mathbf{A}}^{(b)}_{2:L,:},\ \tilde a_t^{(b)}\big)$
  \EndFor
  \State \Return $\{\mathbf{R}^{(b)}\}_{b=1}^B$
\EndFunction
\For{$i=1{:}I$}
  \State \textbf{Sample \& clamp.} $\tilde{\mathbf{a}}^{(b)} \sim \mathcal{N}(\bm\mu,\mathrm{diag}(\bm\sigma^2))$ for $b=1{:}P$;\; clamp each to $[\tilde a_{\text{lo}},\tilde a_{\text{hi}}]$
  \State \textbf{Score.} $\mathbf{R}^{(b)} \gets \Call{Rollout}{\{\tilde{\mathbf{a}}^{(b)}\}}$
  \State \textbf{Elites \& update.} Select top-$K$ elites by $\mathbf{R}^{(b)}$; set $\bm\mu\gets \text{mean}(\text{elites})$;\; $\bm\sigma\gets \text{std}(\text{elites})+\varepsilon$ with $\varepsilon=10^{-6}$
\EndFor
\State \textbf{Return in original units.} $\tilde a_1^\star \gets \bm\mu_{1}$;\; $a_t^\star \gets \mu_{x,i_a} + \sigma_{x,i_a}\cdot \tilde a_1^\star$
\State \textbf{return} $a_t^\star$
\end{algorithmic}
\end{algorithm}
\subsection{MPC/CEM Planner}
\label{subsec:wms3m-planning-compact}
Because planning operates on the same standardized pipeline as inference, we first standardize the current window and map the empirical PRB bounds to scaled coordinates to avoid out-of-distribution proposals (Alg.~\ref{alg:wms3m_plan_cem}, lines~1--3; see also \S\ref{subsec:wms3m-setup}). All sampling, clamping, rewards, and rollouts are then performed in standardized space, with only the selected action mapped back to original units at the end (Alg.~\ref{alg:wms3m_plan_cem}, line~4; see also \S\ref{subsec:wms3m-infer}). We initialize a Gaussian search distribution over the $H$-step PRB sequence with mean at the last observed PRB and a moderate standard deviation, and set the elite count $K=\max(1,\lfloor \rho P\rfloor)$ (Alg.~\ref{alg:wms3m_plan_cem}, lines~5--7). The \textsc{Rollout} routine clones the window and action history and, for $t=1{:}H$, computes the deterministic summary $\bd$ with the causal MS$^{3}$M stack and uses the \emph{prior mean} latent for efficiency (Alg.~\ref{alg:wms3m_plan_cem}, lines~10--11), decodes the next KPI frame with $\Dec_{\text{full}}$ (Alg.~\ref{alg:wms3m_plan_cem}, line~12; see also \S\ref{subsec:wms3m-world}), applies the candidate action $\tilde a_t$ (Alg.~\ref{alg:wms3m_plan_cem}, line~13), evaluates the shaped reward over \{SINR, SE, BLER, Delay, PRB, RSRP\} in scaled space (Alg.~\ref{alg:wms3m_plan_cem}, lines~14--17), adds a smoothness penalty $\lambda_{\mathrm{sm}}\lvert \tilde a_t-\tilde a_{t-1}\rvert$ (Alg.~\ref{alg:wms3m_plan_cem}, line~18), accumulates return, and advances the window by appending the decoded frame and action (Alg.~\ref{alg:wms3m_plan_cem}, lines~19--21),

\begingroup
\footnotesize
\begin{equation}
\label{eq:step-reward}
\begin{aligned}
r_{t+h}
&= \underbrace{w_{\mathrm{SINR}}\;\widehat{y}_{t+h}[i_{\mathrm{SINR}}]}_{\text{higher better}}
 + \underbrace{w_{\mathrm{SE}}\;\widehat{y}_{t+h}[i_{\mathrm{SE}}]}_{\text{higher better}}
 + \underbrace{w_{\mathrm{RSRP}}\;\widehat{y}_{t+h}[i_{\mathrm{RSRP}}]}_{\text{higher better}} \\
&\quad
 - \underbrace{w_{\mathrm{BLER}}\;\widehat{y}_{t+h}[i_{\mathrm{BLER}}]}_{\text{lower better}}
 - \underbrace{w_{\mathrm{Delay}}\;\widehat{y}_{t+h}[i_{\mathrm{Delay}}]}_{\text{lower better}}
 - \underbrace{w_{\mathrm{PRB}}\;\tilde u_{t+h}}_{\text{resource cost}} \\
&\quad
 - \underbrace{\lambda_{\mathrm{sm}}\;\big|\tilde u_{t+h}-\tilde u_{t+h-1}\big|}_{\text{smoothness penalty}},
\end{aligned}
\end{equation}
\endgroup
with nonnegative weights ($w$) and $\lambda_{\mathrm{sm}}\!\ge\!0$. Finally, returning batch returns (Alg.~\ref{alg:wms3m_plan_cem}, line~22). CEM then samples a population, clamps to $[\tilde a_{\mathrm{lo}},\tilde a_{\mathrm{hi}}]$, scores via \textsc{Rollout}, and updates the Gaussian using the top-$K$ elites (Alg.~\ref{alg:wms3m_plan_cem}, lines~23--26). The first element of the final elite mean is de-standardized to produce the next PRB recommendation $a_t^\star$ (Alg.~\ref{alg:wms3m_plan_cem}, lines~27--28).

\section{Experimental Setup}
\label{sec:experimental-setup}

\subsection{Data Collection}
\label{subsec:Data-Collection}
The data-collection steps and all related implementation details are discussed comprehensively in our previous works~\cite{Rezazadeh2025RivalingTM, Rezazadeh2025ReservoirAugmented, dai2025orankpi}. The dataset\footnote{A curated release is hosted on IEEE DataPort: \url{https://ieee-dataport.org/documents/video-streaming-network-kpis-o-ran-testing}} is publicly available to support reproducibility and enable follow\mbox{-}on research.

\subsection{KPIs and Preprocessing}
Each sample consists of a lookback window of length \(L\) with feature vector \(\bx_t\in\mathbb{R}^{F}\) that includes modulation/coding indicators and radio-quality metrics (e.g., modulation and coding scheme (MCS), channel quality indicator (CQI), rank indicator (RI), precoder matrix indicator (PMI), reference signal received quality (RSRQ), RSRP, received signal strength indicator (RSSI), and SINR), traffic/load indicators (e.g., buffer occupancy, SE, BLER, and delay), and the control variable used for planning (PRBs). Targets \(\by_{t+1}\) are one-step-ahead KPIs; we report the univariate RSRP head and also train a full multivariate reconstruction head for representation learning. Standardization is per feature using mean and standard deviation computed on the training split only; the same affine maps are then frozen and applied to validation and test periods. Windows and labels are formed by sliding in time without overlap between the train/validation/test ranges. To remain in-distribution during planning, admissible PRB values are clipped to the empirical \([5,95]\)th percentiles measured on the training set and mapped to the model's standardized space.

\subsection{Model Hyperparameters}
The world model couples a causal multi-scale SSM front end with a compact stochastic latent. We use the following configuration: embedding width \(d{=}192\), SSM mixture layers \(L_\ell{=}4\), per-channel HiPPO–LegS state size \(N{=}64\), mixture components \(M{=}4\) with learned positive time steps (bilinear discretization), latent dimension \(d_z{=}48\), dropout \(0.1\), and lookback length \(L\) matching the recorded window. The target head is heteroscedastic (mean\(+\)log-variance clamped to \([-8,8]\)); a small autoregressive skip from the last input is combined through a bounded gain \(\tanh(\kappa)\). Training uses AdamW with learning rate \(2\!\times\!10^{-3}\), weight decay \(10^{-4}\), batch size \(256\), cosine warm restarts stepped per mini-batch, gradient clipping at norm \(1.0\), and patience-based early stopping. We employ KL annealing from \(0.01\) to \(1.0\) over the first 20 epochs and scheduled mixing between posterior and prior encoders from \(1.0\) to \(0.5\) on the same schedule. Light input Gaussian noise (std \(0.01\)) and feature-channel dropout (\(0.1\)) improve robustness. At test time, we average \(S{=}8\) prior samples for the target prediction. For MPC/CEM planning, we set horizon \(H{=}8\), population \(256\), elite fraction \(0.1\), four CEM iterations, an action-smoothness penalty \(0.05\), and reward weights in standardized space that positively score SINR/SE/RSRP and penalize BLER/Delay/PRB usage.

To ensure fair comparisons, all neural baselines are matched in adequate capacity and trained with the same optimizer, early stopping, and chronological data splits. Inputs at time $t$ contain only measurements available up to (and including) $t$, targets are at $t{+}1$, and standardization is fit on the training split only and frozen thereafter. Errors are reported in original physical units after inverse transforms; we additionally report latency per forward pass to reflect Near-RT feasibility. This protocol isolates architectural differences from data leakage and tuning effects, making the comparison both reproducible and operationally meaningful.

\subsection{Baselines}
We evaluate the proposed world-modeling approach against baselines spanning state-space, attention, and recurrence inductive biases under leakage-safe, strictly causal evaluation. Concretely, we include: (a) a strictly multi-scale structured state-space mixture (MS$^{3}$M \cite{Rezazadeh2025RivalingTM}; SSM/HiPPO–LegS depthwise kernels with light channel mixing) that shares our causal front end but omits the stochastic latent and agentic rollouts—serving as a strong “pure SSM” reference isolating the value of compact latents for calibration and counterfactuals; (b) the recurrent–attention hybrid RWKV \cite{peng2023rwkv}, which fuses token-by-token recurrence with attention-style mixing and thus stresses exposure-bias behavior under one-step-ahead protocols; (c) a linear-attention Transformer (Performers \cite{choromanski2021rethinking}) that approximates softmax attention with kernel features to retain global content-based interactions at near-linear cost, suitable for near–real-time regimes (e.g., O\!-RAN telemetry); (d) a retentive state-space network (RetNet \cite{sun2023retentive}) offering long-memory state updates and linear-time inference as a scalable SSM-style comparator; and (e) tokenized sequence models in the style of Chronos (Chronos-T5 / Chronos-GPT \cite{ansari2024chronos}), used here as decoder/seq2seq baselines trained from scratch on our dataset to probe the value of strong sequence priors even without unit-aware calibration. 

\begin{mdframed}[backgroundcolor=gray!5,linecolor=gray!50]
\footnotesize
To ensure fair timing/memory budgets, we use unified data processing and training for all models and implement compact baseline variants that preserve each method's core inductive bias. These are not exact reproductions of the original codebases; the goal is comparability under a common setup rather than paper-by-paper replication.
\end{mdframed}
\begin{table*}[t!]
\centering
\scriptsize
\setlength{\tabcolsep}{8pt}
\caption{Benchmark Results for Proposed WM-MS$^{3}$M vs.\ Baselines.}
\label{tab:results-ms3m}
\resizebox{\textwidth}{!}{
\begin{tabular}{lrrrrrrrr}
\toprule
\textbf{Model} & \textbf{RMSE} & \textbf{MAE} & \textbf{MSE} & \textbf{Skill (R)} & \textbf{Skill (M)} & \boldmath$R^2$ & \textbf{\#Params} & \textbf{Infer (s)} \\
\midrule
\text{WM--MS}$^{3}$\text{M} & 0.291687 & 0.167417 & 0.085081 & 0.918448 & 0.938430 & 0.993035 & 477{,}802 & 0.000647 \\
\midrule
MS$^{3}$M \cite{Rezazadeh2025RivalingTM}     & 0.291806 & 0.170298 & 0.085151 & 0.918414 & 0.937370 & 0.993030 & 698{,}449     & 0.000643 \\
RWKV \cite{peng2023rwkv}          & 1.480259 & 0.407525 & 2.191166 & 0.576394 & 0.848496 & 0.812361 & 26{,}269{,}696 & 0.002103 \\
Performers \cite{choromanski2021rethinking}     & 0.786704 & 0.276759 & 0.618902 & 0.774869 & 0.897110 & 0.947001 & 14{,}156{,}160 & 0.002634 \\
RetNet \cite{sun2023retentive}        & 0.598374 & 0.208329 & 0.358051 & 0.828763 & 0.922550 & 0.969338 & 18{,}930{,}432 & 0.001522 \\
Chronos-T5 \cite{ansari2024chronos}     & 2.518750 & 1.673298 & 6.344100 & 0.279208 & 0.377925 & 0.456727 & 14{,}209{,}536 & 0.000790 \\
Chronos-GPT \cite{ansari2024chronos}    & 0.445510 & 0.183071 & 0.198479 & 0.872508 & 0.931941 & 0.983003 & 6{,}840{,}320  & 0.000749 \\
\bottomrule
\end{tabular}%
}
\\[2pt]
\raggedright\scriptsize \textit{Notes:} Inference time is per-sample wall-clock seconds.%
\end{table*}
\section{Performance and Numerical Results}
\label{sec:perf-numerical}

\subsection{Accuracy Against the Primary SSM Baseline}
As shown in \cref{tab:results-ms3m}, compared to its counterpart (MS\(^3\)M), WM--MS\(^3\)M yields consistent accuracy gains while using substantially fewer parameters.
In addition to absolute error metrics, performance is assessed using normalized skill scores computed relative to a reference baseline; Skill~(R), which measures relative improvement based on RMSE, and Skill~(M), which measures improvement based on MSE, with higher values indicating better performance~\cite{Hyndman2006Foresight, Murphy1988Skill}.
RMSE decreases from \(0.291806\) to \(0.291687\) (about \(\SI{0.04}{\percent}\)), MAE drops from \(0.170298\) to \(0.167417\) (about \(\SI{1.69}{\percent}\)), and MSE improves from \(0.085151\) to \(0.085081\) (about \(\SI{0.08}{\percent}\)).
Skill~(R) rises from \(0.918414\) to \(0.918448\) and Skill~(M) from \(0.937370\) to \(0.938430\), with \(R^2\) nudging from \(0.993030\) to \(0.993035\).
Crucially, parameters fall from \(698{,}449\) to \(477{,}802\), a reduction of roughly \(\SI{31.59}{\percent}\), while the inference time remains effectively unchanged (\(0.000643\) s vs.\ \(0.000647\) s).
These results indicate that the compact stochastic latent and dual-decoder design confer measurable benefits in accuracy and calibration over the same causal SSM front end, without compromising latency and while improving model efficiency.

\subsection{Gains Over Attention/Hybrid and Pretrained Baselines}

\subsubsection{Performance (Accuracy)}
WM--MS\(^3\)M achieves the strongest overall error metrics in \cref{tab:results-ms3m}, with clear improvements in RMSE and MAE over attention, hybrid, and pretrained baselines (e.g., RetNet, Performers, RWKV, and Chronos-GPT). These gains stem from its causal multi-scale SSM front end, compact stochastic latent for calibrated uncertainty, and dual-decoder design that focuses capacity on the operational KPI while maintaining strict causality. In practice, this combination yields lower nowcasting error under the same leakage-safe evaluation protocol.

\subsubsection{Latency (Near-real-time Deployability)}
Despite higher modeling fidelity, WM--MS\(^3\)M remains in the fastest latency tier (effectively on par with MS\(^3\)M) and is notably quicker than attention and hybrid alternatives. Because we use depthwise, causal SSM filters, the compute scales roughly linearly with the number of channels and the receptive-field length, rather than quadratically with token-to-token interactions as in full self-attention. Even linear-attention variants still rely on large, dense projections, which increase inference time. Consequently, WM--MS\(^3\)M offers lower wall-clock latency per sample—well-suited for Near-RT O\!-RAN loops.
\begin{mdframed}[
  backgroundcolor=gray!5,
  linecolor=gray!50,
  linewidth=0.5pt,
  roundcorner=2pt,
  innerleftmargin=8pt,
  innerrightmargin=8pt,
  innertopmargin=6pt,
  innerbottommargin=6pt,
  skipabove=6pt,
  skipbelow=6pt,
  userdefinedwidth=\columnwidth,
  align=center
]
\centering
\footnotesize
\text{Inference Latency Ranking (fastest $\to$ slowest)}\\[2pt]
\(
  \text{MS$^{3}$M} \;\approx\; \textbf{WM--MS$^{3}$M}
  \;\lesssim\;
  \text{Chronos-GPT}
  \;\lesssim\;
  \text{Chronos-T5}
  \;\lesssim\;
  \text{RetNet}
  \;\lesssim\;
  \text{RWKV}
  \;\lesssim\;
  \text{Performers}
\)\\[3pt]
\end{mdframed}

\subsubsection{Complexity Summary}
Parameter count is a practical proxy for model capacity. It directly influences memory footprint and the number of multiply–accumulate operations, so \emph{more parameters generally imply higher computational and memory cost}. However, parameters alone do not fully determine runtime complexity; the \emph{operator structure} is equally essential. \text{WM--MS\textsuperscript{3}M} is compact (fewer parameters than attention/hybrid baselines). It relies on structured state-space layers whose per-step cost scales favorably, whereas attention mechanisms introduce quadratic (or large linear) costs with respect to context length and hidden width.

Let $L$ be the SSM history window and let $T \!=\! 2L$ be the tokenized input length used by the token language models (LMs) (Chronos-GPT, Chronos-T5, RetNet, Performers, RWKV), where we alternate \texttt{RSRP}/\texttt{PRB} tokens. Let $d$ denote the model width, $h$ the number of heads with $d_h \!=\! d/h$, $C$ the number of SSM mixture components, $N$ the SSM state size, $m$ the Performers random features, and $(I, P, H)$ the CEM iterations, population size, and horizon for planning.

\begin{table}[ht]
\centering
\scriptsize
\setlength{\tabcolsep}{6pt}
\caption{Asymptotic Time Complexities.}
\label{tab:complexity-time}
\begin{tabularx}{\linewidth}{l l l}
\toprule
\textbf{Method} & \textbf{Train time (per layer)} & \textbf{Inference time (1 step)} \\
\midrule
\textbf{WM--MS\textsuperscript{3}M} & $O(d\,L^2)$ & $O(d\,L^2)$ \\
\quad + CEM planning & \multicolumn{2}{l}{$\;\;$Adds $O(IPH \cdot d\,L^2)$ (rollouts)} \\
\text{MS\textsuperscript{3}M} & $O(d\,L^2)$ & $O(d\,L^2)$ \\
\text{RWKV} & $O(T d)$ & $O(T d)$ \\
\text{Performers} & $O(T d\,m)$ & $O(T d\,m)$ \\
\text{RetNet} & $O(T d)$ & $O(T d)$ \\
\text{Chronos-T5} & $O(T^2 d)$ (enc) $+\;O(T d)$ (dec) & $O(T d)$ (single target) \\
\text{Chronos-GPT} & $O(T^2 d)$ & $O(T d)$ (KV cache) \\
\midrule
\multicolumn{3}{l}{\textit{Symbols:} $L$ history window; $T{=}2L$ token length; $d$ width; $m$ random features;}\\
\multicolumn{3}{l}{\hspace*{1.5em}$(I,P,H)$ CEM iters/pop/horizon. $C,N$ absorbed into constants.}
\\
\bottomrule
\end{tabularx}
\end{table}

The SSM forecasters (\text{MS\textsuperscript{3}M} and the backbone of \text{WM--MS\textsuperscript{3}M}) perform depthwise, time-domain SSM convolutions of length $L$, giving per-layer time complexity of $O(d\,L^2)$. \text{WM--MS\textsuperscript{3}M} adds stochastic latent inference and decoders with cost complexity of $O(d)$ (negligible in $L$), and when used with MPC/CEM planning it rolls the model $I\times P\times H$ times; thus planning contributes an additional time complexity of $O(IPH \cdot d\,L^2)$ on top of a single forward pass (see Table~\ref{tab:complexity-time}). For token LMs, \text{Chronos-GPT} (decoder-only attention) cost complexity is $O(T^2 d)$ per layer during training, and $O(T d)$ per decoded token. \text{Chronos-T5} has the complexity of $O(T^2 d)$ in the encoder plus $O(T d)$ for the (single-token) decoder step that cross-attends to the encoder. \text{RetNet} and \text{RWKV} use retention/recurrent updates with linear sequence scaling, giving the time complexity of $O(T d)$. \text{Performers} replaces quadratic attention with FAVOR$^{+}$ random features \cite{choromanski2021rethinking}, giving the time complexity of $O(T d\,m)$.

\subsection{Qualitative Visualization and Error Analysis}
\label{subsec:qual-viz}

Figure~\ref{fig:gt-pred-kpis} visualizes the last 500 test points for four key KPIs with ground truth (solid) and one-step predictions (dashed) stacked vertically in the order (a) RSRP, (b) SINR, (c) CQI, and (d) PRBs. Curves are shown in original physical units to facilitate a direct, unit-aware assessment of fidelity and bias. This qualitative view complements the aggregate metrics in Table~\ref{tab:results-ms3m}.

\subsubsection{Trend Fidelity and Short-horizon Dynamics}
Across Figure~\ref{fig:gt-pred-kpis} (a)--(c), the model tracks slow trends and most short-horizon fluctuations with small residuals, indicating that the causal multi-scale SSM backbone and compact latent capture the dominant temporal structure without violating strict causality. Occasional under-/overshoots correspond to abrupt regime changes where any strictly one-step causal forecaster is maximally stressed.

\subsubsection{Cross-feature Consistency}
\begin{figure}[t!]
\centering
\includegraphics[width=\linewidth]{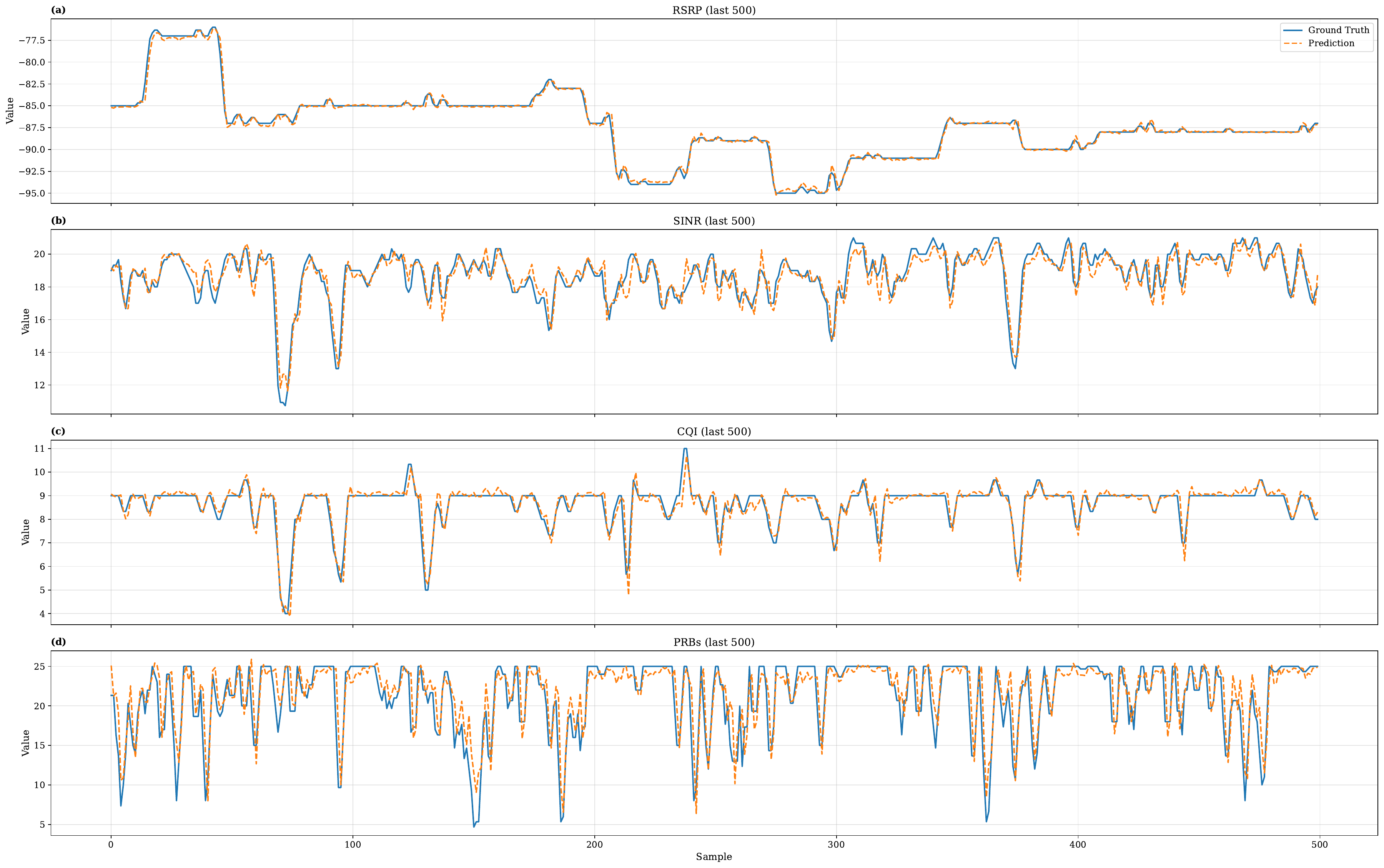}
\caption{Last 500 test samples: ground truth (\emph{solid}) vs.\ one-step model prediction (\emph{dashed}) for four KPIs—(a) RSRP, (b) SINR, (c) CQI, and (d) PRBs. Curves are shown in original units.}
\label{fig:gt-pred-kpis}
\end{figure}

Figure~\ref{fig:gt-pred-kpis} (d) shows that PRBs---although used as the control channel during planning---are also reconstructed accurately by the multivariate head. This internal consistency across KPIs is important operationally, as the world model leverages cross-feature couplings (e.g., PRBs$\leftrightarrow$SINR/RSRP) when rolling out counterfactuals and for short-horizon planning.

\subsubsection{Error Localization}
The most significant deviations appear near sharp transitions (e.g., scheduler or load shifts). Away from such events, residuals remain low and visually stationary, aligning with the quantitative improvements in RMSE/MAE/MSE reported for WM--MS$^{3}$M in Table~\ref{tab:results-ms3m}. In practice, these localized errors are the natural consequence of strict leakage-safe evaluation. They can be further mitigated by increasing the lookback, modestly widening the mixture of time scales, or ensembling prior samples at test time.

\subsection{Counterfactual ``What--If'' Forecasting and Short--Horizon Planning}
\label{subsec:whatif}

We frame evaluation as a \emph{what--if} study. Given the same recent KPI window $\mathbf{X}_{t-L+1:t}$, we ask what the next $H$ steps of KPIs would look like under alternative future PRB controls. For a candidate sequence $u_{t:t+H-1}$, the world model returns a counterfactual trajectory $\mathbf{y}_{t+1:t+H}(u)=\mathrm{WM}\!\big(\mathbf{X}_{t-L+1:t},u_{t:t+H-1}\big)$ and a shaped return $R(u)$ (scaled space) that trades SINR/SE/RSRP against BLER/Delay and PRB cost, with an action\hyp{}smoothness penalty. The context window is fixed; only the control path changes.

\begin{figure}[t!]
  \centering
  \includegraphics[width=\linewidth]{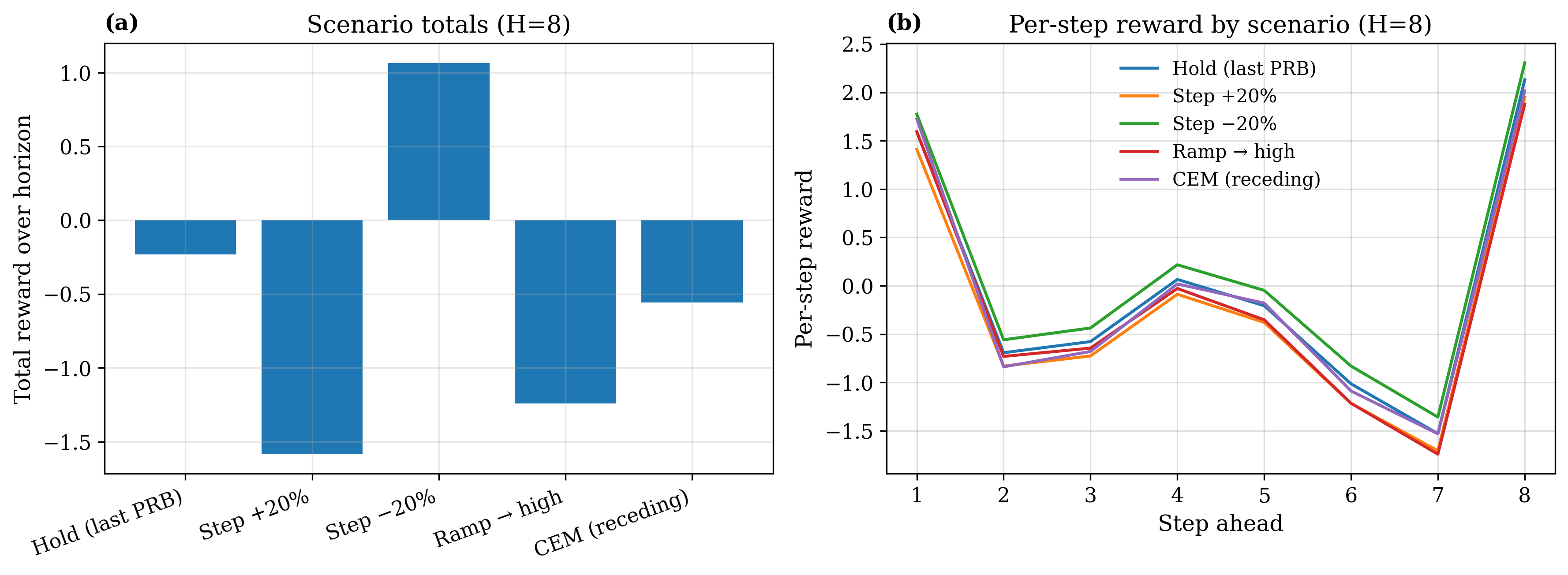}
  \caption{Reward comparison across PRB control scenarios over horizon $H\!=\!8$. \textbf{(a)} Total cumulative reward per scenario (higher is better). \textbf{(b)} Per\hyp{}step reward trajectories showing when gains/losses occur. 
  }
  \label{fig:whatif-panels}
\end{figure}

We test five interventions over $H\!=\!8$: (i) \emph{Hold (last PRB)}, (ii) \emph{Step $+20\%$} (immediate increase then hold), (iii) \emph{Step $-20\%$} (immediate decrease then hold), (iv) \emph{Ramp$\to$high} (linear increase to the empirical high bound), and (v) a receding\hyp{}horizon \emph{CEM} policy using the same reward and data\hyp{}driven bounds. Figure~\ref{fig:whatif-panels}\,(a) agrees with Table~\ref{tab:whatif}: \textbf{Step $-20\%$} achieves the highest total reward ($+1.066$), \emph{Hold} is slightly negative ($-0.232$), and PRB\hyp{}increasing scripts (\emph{Step $+20\%$}, \emph{Ramp$\to$high}) are worst (respectively $-1.585$ and $-1.242$) due to PRB cost and a mild BLER penalty despite slight SINR upticks. The receding\hyp{}horizon \emph{CEM} stays close to \emph{Hold} ($-0.501$), reflecting the short horizon, smoothness penalty, and clamped action range.

Figure~\ref{fig:whatif-panels}\,(b) shows where gains/losses occur across the horizon. All scenarios start with a positive reward at $h{=}1$ (about $1.4$–$1.8$), dip at $h{=}2$ (roughly $-0.6$ to $-0.8$), recover modestly through $h{=}4$–$5$, reach their trough near $h{=}7$ (about $-1.4$ to $-1.8$), and rebound strongly at $h{=}8$ (about $2.0$–$2.3$). The \emph{Step $-20\%$} curve is consistently best (or tied) over most steps, explaining its superior total. In short, under the configured reward and on this window, \emph{reducing} PRBs by $20\%$ offers the best trade-off between KPI quality and control cost over $H{=}8$.

\subsection{Why WM--MS\texorpdfstring{\(^3\)}{}M Outperforms}
In our results (Table~\ref{tab:results-ms3m}, Figure~\ref{fig:whatif-panels}), each improvement maps to a concrete mechanism. The \emph{heteroscedastic target head} plus a \emph{compact latent} (with KL annealing and posterior--prior mixing) reduces typical errors without over-smoothing rare spikes, thus \text{MAE decreases by 1.69\%} while \text{RMSE remains essentially unchanged} versus MS$^{3}$M. The \emph{dual-decoder split} (full-frame decoder + KPI-focused head with a bounded AR skip) together with a \emph{depthwise, strictly causal SSM} backbone targets capacity where it matters, yielding \text{$\sim$32\% fewer parameters} at the \text{same latency}. The \emph{telemetry-matched SSM inductive bias} and a leakage-safe pipeline (chronological splits, train-only scalers) account for the \text{35--80\% RMSE gains} over attention/hybrid baselines while being \text{2.3--4.1$\times$ faster}. In what-if planning, the model’s learned diminishing SINR/RSRP returns for higher PRBs, combined with the reward’s PRB/BLER/Delay penalties (Equation~\eqref{eq:step-reward}) and in-distribution action bounds, make \text{Step $-20\%$ PRB} optimal on the evaluated window—explaining why WM--MS$^{3}$M yields the specific outcomes we report.

\begin{mdframed}[
  backgroundcolor=gray!5,
  linecolor=gray!50,
  linewidth=0.5pt,
  roundcorner=2pt,
  innerleftmargin=8pt,
  innerrightmargin=8pt,
  innertopmargin=6pt,
  innerbottommargin=6pt,
  skipabove=6pt,
  skipbelow=6pt,
  userdefinedwidth=\columnwidth,
  align=center
]
\centering
\footnotesize
\text{Accuracy–Efficiency Ranking (best $\to$ worse)}\\[2pt]
\(
  \textbf{WM--MS$^{3}$M}
  \;\lesssim\;
  \text{MS$^{3}$M}
  \;\lesssim\;
  \text{RetNet}
  \;\lesssim\;
  \text{Performers}
  \;\lesssim\;
  \text{RWKV}
  \;\lesssim\;
  \text{Chronos-GPT}
  \approx
  \text{Chronos-T5}
\)\\[3pt]
\scriptsize Jointly considers lower errors, lower latency, and fewer parameters (see Table~\ref{tab:results-ms3m}).
\end{mdframed}
\subsection{Operational Implications}
Although the absolute error improvements over MS\(^3\)M are modest, they are achieved with approximately \(\SI{31.6}{\percent}\) fewer parameters and near-identical latency, which is valuable for edge-side deployment in Near-RT control loops. The large margins over attention, hybrid, and pretrained sequence models underscore that a causal, SSM-first inductive bias, combined with a calibrated latent, better matches O\!-RAN telemetry than token-centric or global-attention architectures. Beyond accuracy, the same substrate enables uncertainty-aware counterfactual analysis and short-horizon MPC/CEM planning without retraining, providing auditable decision support that extends past next-step prediction and aligns with safety and transparency requirements in 6G O\!-RAN.
\begin{table}[t!]
  \centering
  \caption{What--if Forecasting over $H\!=\!8$.}
  \label{tab:whatif}
  \small
  \setlength{\tabcolsep}{5pt}
  \resizebox{\columnwidth}{!}{%
  \begin{tabular}{lrrrr}
    \toprule
    \textbf{Scenario} & \textbf{Reward} & \textbf{avg RSRP} & \textbf{avg SINR} & \textbf{avg BLER}\\
    \midrule
    Hold (last PRB)   & $-0.232$ & $-87.423$ & $18.450$ & $5.155$ \\
    Step +20\%        & $-1.585$ & $-87.427$ & $18.474$ & $5.224$ \\
    Step --20\%       & $\mathbf{+1.066}$ & $-87.417$ & $18.417$ & $5.109$ \\
    Ramp $\to$ high   & $-1.242$ & $-87.428$ & $18.467$ & $5.221$ \\
    CEM (receding)    & $-0.501$ & $-87.424$ & $18.450$ & $5.152$ \\
    \bottomrule
  \end{tabular}%
  }
\end{table}
\section{Conclusion}
\label{sec:conclusion}
In this paper, we have proposed an \emph{agentic world modeling} paradigm for 6G O\text{-}RAN that treats actions as first-class causes and plans over \emph{consequences} under calibrated uncertainty. We have applied this approach with a PRB-conditioned \emph{WM--MS\(^3\)M} architecture that couples a strictly causal, multi-scale, and structured state-space front end with compact stochastic latent and dual decoders. The design enforces leakage-safe training, supports counterfactual (what-if) forecasting, and enables short-horizon MPC/CEM planning inside data-driven operating range. On realistic O\text{-}RAN traces, WM--MS\(^3\)M matched or surpassed strong sequence baselines while using fewer parameters and maintaining Near-RT latency. Beyond point accuracy, the model provides uncertainty summaries and auditable rollouts that better align with the safety and transparency requirements of radio control than LLM-centric approaches. Overall, the results indicate that world models with calibrated uncertainty can deliver \emph{control-grade prediction} and decision support in O\text{-}RAN, while reserving LLMs for complementary roles (e.g., orchestration, explanation, policy templating) under strict grounding. We outline several directions to extend this study:
\begin{itemize}[leftmargin=*,nosep]
\item \textbf{Closed-loop field trials.} Integrate the predictive and planning APIs with Near-RT RIC xApps/rApps and evaluate end-to-end on live O-RAN segments, including safety guards, rollback, and human-in-the-loop adjudication.
\item \textbf{Richer controls and multi-cell coordination:} Expand the action channel beyond PRBs (power, MCS, beams, timers) and model coordinated policies across cells/sectors with multi-agent rollouts and interference-aware rewards.
\item \textbf{Long-horizon and hierarchical planning:} Combine short-horizon MPC with value-function critics or options for hierarchical control; study horizon sensitivity and stability under receding-horizon execution.
\item \textbf{Adaptation and robustness:} Add online/federated updates, change-point detection, and domain shift defenses (e.g., priors, adapters, ensembling) to maintain calibration under non-stationarity and vendor diversity.
\item \textbf{Causal structure and interpretability:} Incorporate causal regularizers/latent interventions; expose sensitivities and counterfactual explanations at KPI and action levels for auditability and SLA diagnostics.
\end{itemize}
Pursuing these directions will move world-model-based control from accurate prediction and what-if analysis to \emph{safe, auditable, and scalable} automation in open 6G networks.

\bibliographystyle{IEEEtran}
\bibliography{refs}

\end{document}